%% file: main.tex
\documentclass[aps,prd,reprint,superscriptaddress,nofootinbib,floatfix]{revtex4-2}

\usepackage{amsmath,amssymb,bm,booktabs,xcolor,graphicx}
\usepackage[normalem]{ulem}
\usepackage[
  colorlinks=true,
  linkcolor=blue!55!black,
  citecolor=blue!55!black,
  urlcolor=blue!55!black,
  pdfborder={0 0 0},
  pdftitle={Mixing-suppressed inelastic dark matter: a minimal model for the LZ 248 keV event},
  pdfauthor={Seung J. Lee and Taewook Youn}
]{hyperref}

\input{SD_LZ_macros}

\makeatletter\expandafter\def\csname b@apsrev42Control\endcsname{}\makeatother
\begin{document}

\title{Mixing-suppressed inelastic dark matter: a minimal model for the LZ 248~keV event}

\author{Seung~J.~Lee}
\email{sjjlee@kias.re.kr}
\affiliation{School of Physics, KIAS, 85 Hoegi-ro, Dongdaemun-gu, Seoul 02455, Korea}
\affiliation{Quantum Universe Center, KIAS, 85 Hoegi-ro, Dongdaemun-gu, Seoul 02455, Korea}
\author{Taewook~Youn}
\email{taewookyoun@fas.harvard.edu}
\affiliation{School of Physics, KIAS, 85 Hoegi-ro, Dongdaemun-gu, Seoul 02455, Korea}
\affiliation{Gravity, Spacetime, and Particle Physics (GRASP) Initiative, Harvard University,
17 Oxford Street, Cambridge, MA 02138, USA}

\date{\today}

\begin{abstract}
We construct a minimal Majorana singlet--vector-like-doublet model for the single
nuclear-recoil-like event reported by LZ near $248\keV$. At the splitting inferred from the recoil
energy, an electroweak-strength $Z$ transition predicts thousands of events; singlet--doublet
mixing suppresses the rate without moving the recoil spectrum. Fixed-coupling interpretations
instead require a larger splitting in the extreme halo tail, but solar gravitational acceleration
removes this suppression, while the larger splitting shifts recoils toward LZ's empty high-energy
sideband. In our model
the same mixing suppresses solar capture and annihilation, while the small mass gap required for
coannihilation weakens Higgs-mediated cooling, allowing the captured population to remain extended
and out of equilibrium. Imposing the relic abundance and normalizing the rate to one event leaves
a two-dimensional mass--splitting parameter space. A thermalized population gives the
conservative IceCube limit $\delta=302$--$305\keV$; with our nonthermal cooling ansatz and elastic
scattering treated at tree level, the nominal limit is $\delta=331$--$343\keV$ near the
$730$--$733\GeV$ relic-density endpoint, which is also favored by the recoil spectrum. We outline
a gauged $U(1)_N$ origin for the parity and splitting and a candidate $R$-symmetric supersymmetric
embedding.
\end{abstract}

\maketitle
\raggedbottom

\input{SD_LZ_body}
\nocite{apsrev42Control}
\bibliographystyle{apsrev4-2}
\bibliography{refs}

\end{document}

%% file: SD_LZ_macros.tex
\newcommand{\keV}{\,\mathrm{keV}}\newcommand{\MeV}{\,\mathrm{MeV}}
\newcommand{\GeV}{\,\mathrm{GeV}}\newcommand{\TeV}{\,\mathrm{TeV}}
\newcommand{\cms}{\,\mathrm{cm^3\,s^{-1}}}\newcommand{\cmm}{\,\mathrm{cm^2}}
\newcommand{\sv}{\langle\sigma v\rangle}
\newcommand{\st}{\sin^2\!\theta}\newcommand{\sft}{\sin^4\!\theta}
\newcommand{\Glim}{\Gamma_{\rm lim}}
\newcommand{\Gann}{\Gamma_{\rm ann}}
\newcommand{\PsiN}{\Psi_N}\newcommand{\Psip}{\Psi_\psi}

%% file: SD_LZ_body.tex
\section{Introduction}\label{sec:intro}

LZ has reported one nuclear-recoil-like event at $248\pm23\pm23\keV$ in $2.84$\,t\,yr, with
$0.01$ expected \cite{LZ}. Its local significance is $3.4\sigma$ and its look-elsewhere-corrected
significance is $2.6\sigma$. The absence of accompanying low-energy recoils in a larger exposure
disfavors an elastic interpretation \cite{LZWIMP}. Endothermic scattering
$\chi_1A\to\chi_2A$, with $m_{\chi_2}-m_{\chi_1}=\delta$, removes that low-energy population
\cite{IDM,IDMstatus}. For a representative xenon nucleus with mass number $A=131$ and
$m_{\chi_1}\simeq730\GeV$, the $55\keV$ detector threshold and the kinematic limit evaluated at
the mean Earth speed give lower and upper reference limits of $\delta\simeq243$ and $362\keV$
(Sec.~\ref{sec:thresh})
\cite{Higgsino,DiMauro,SYY,Vis,DP,HCMDM}.

\begin{figure}[tbp]
\centering
\includegraphics[width=\columnwidth]{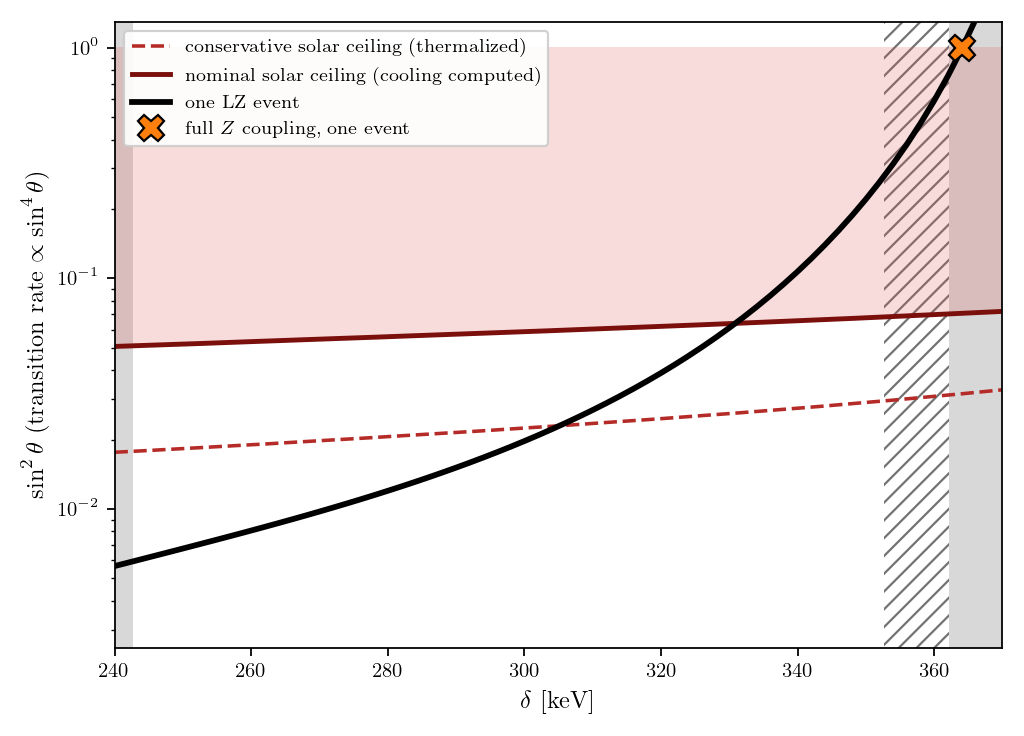}
\caption{Two routes to one LZ event at $m_\chi=730\GeV$. The black curve is fixed by requiring one
expected event in the LZ window; the
cross denotes the transition at full strength, while mixing moves the prediction down the curve. The dashed and solid red
lines show, respectively, the conservative solar ceiling for a thermalized population and the
nominal ceiling obtained with the cooling ansatz; the shaded region above the latter is excluded
under that ansatz. Gray shading marks the reference kinematic limits for xenon
(Sec.~\ref{sec:thresh}), and hatching marks where the expected sideband yield exceeds $0.6$ after
normalizing to one event in the LZ window (Sec.~\ref{sec:shape}).}
\label{fig:money}
\end{figure}

At $m_\chi=730\GeV$ and $\delta\simeq292\keV$, a $Z$ transition of full electroweak strength
would give about $4\times10^3$ events, even though only $0.45\%$ of the standard halo lies above
the inelastic threshold (Sec.~\ref{sec:thresh}). A fixed-coupling interpretation can reduce this
to one event only by raising $\delta$ to about $364\keV$ and selecting an even thinner part of
the halo tail, as in the Higgsino interpretation \cite{Higgsino,WZZ,FanReece}. Singlet--doublet
mixing instead supplies the required suppression through the transition coupling. The rate is
proportional to $\sft$, while the recoil spectrum is unchanged. Thus the two routes differ in whether
the remaining suppression comes from the velocity distribution or the coupling.
Figure~\ref{fig:money} displays this distinction.

Recent work also considers a singlet--doublet model with a doublet-sector splitting
\cite{BorahSahoo2026}, nonthermal Higgsinos or weakly coupled $Z'$ mediators
\cite{Langhoff2026,LeeRandall2026}. Solar studies confirm the strength of the IceCube constraint
\cite{NguyenLindenHooper2026} and its dependence on the annihilation channel and captured population
\cite{DiMauroShaikh2026}.

The difference between the two routes matters in the Sun. Solar gravitational acceleration removes the
reliance on the Galactic high-velocity tail, including the additional suppression produced by a
larger splitting. A fixed-strength interaction therefore leads to efficient capture and, after
equilibration, violates the IceCube bound \cite{PR}. Its larger
splitting also shifts xenon recoils toward the empty $350$--$590\keV$ sideband \cite{RSSX}. The
same mixing suppresses solar capture and gauge-boson annihilation. In addition, the
Majorana $Z$ current is purely off diagonal and provides no elastic cooling, while the small
singlet--doublet mass gap fixed by the relic abundance suppresses Higgs exchange. The captured
population therefore remains extended. Even if it were thermalized, mixing would leave the
reference point with an equilibration factor of only about $0.1$; inefficient cooling suppresses
the annihilation rate further (Sec.~\ref{sec:equil}).

\begin{figure}[tbp]
\centering
\includegraphics[width=\columnwidth]{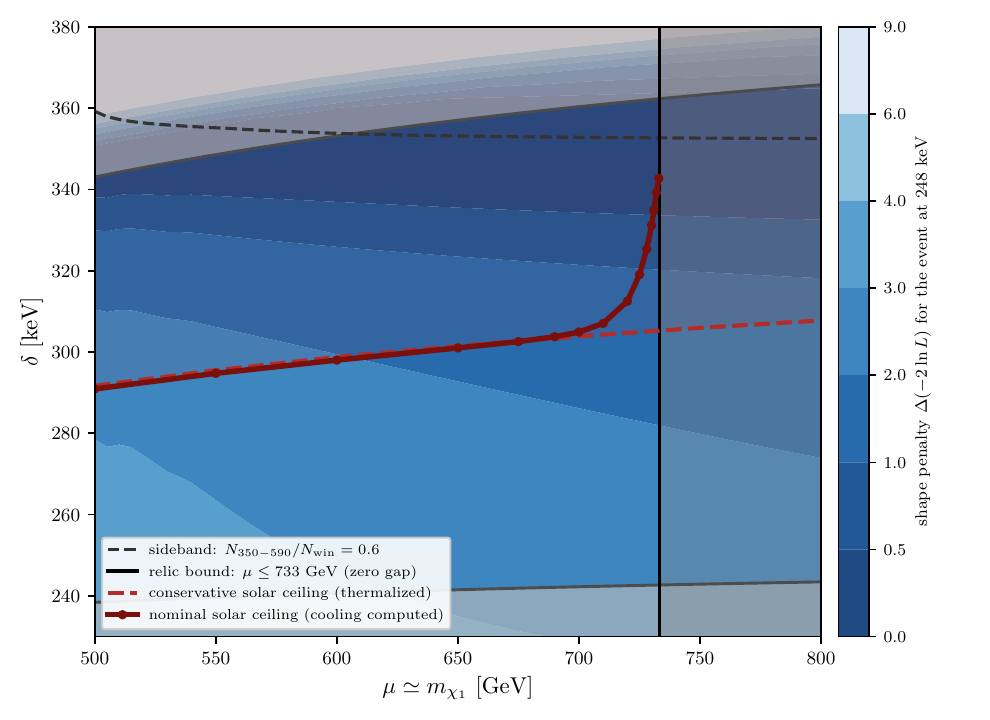}
\caption{Parameter space after fixing $M_D-\mu$ by the relic abundance and $\st$ by one LZ
event. Red curves are the conservative (dashed) and nominal (solid) solar ceilings;
gray shading marks the reference kinematic limits for xenon (Sec.~\ref{sec:thresh}) and the
relic-density endpoint; the gray dashed curve marks a sideband yield of $0.6$ after normalizing to
one event in the LZ window. The color scale gives the likelihood penalty from the recoil shape. The
region below the nominal ceiling, above the lower kinematic limit, and to the left of the
relic-density endpoint is allowed by the constraints considered here.}
\label{fig:plane}
\end{figure}

We realize this mechanism with a Majorana singlet mixed with a vector-like electroweak doublet.
The relic abundance fixes $M_D-\mu$, and one LZ event fixes the mixing, leaving the mass and
splitting to be tested by the recoil spectrum and the Sun. The recoil spectrum favors
$\delta=331$--$343\keV$ near the $730$--$733\GeV$ relic-density endpoint, where the nominal solar
limit is saturated; a deliberately conservative thermalized population instead gives
$\delta=302$--$305\keV$. Figure~\ref{fig:money} shows the suppression mechanism, while
Fig.~\ref{fig:plane} gives the parameter space after imposing constraints from the relic density,
recoil spectrum, sideband, and solar neutrino limit.

The recoil spectrum favors larger splittings, so the best-fitting allowed points saturate the
nominal solar limit. Their shape penalty is a goodness-of-fit statement and does not revise
LZ's discovery significance. Points below the nominal ceiling remain
allowed, and further recoils test the model directly: several events above $240\keV$ would make
the recoil spectrum favor splittings above the solar limit.

This paper is organized as follows.
Section~\ref{sec:model} presents the model, Sec.~\ref{sec:calc} the constraints and results,
Sec.~\ref{sec:uv} possible ultraviolet completions, and Sec.~\ref{sec:fate} our conclusions.

\section{The model}\label{sec:model}

\subsection{Fields, symmetry and Lagrangian}\label{sec:fields}\label{sec:sym}
We add four Weyl fermions (Table~\ref{tab:fields}), all odd under the $Z_2$ that stabilizes the
dark matter. In the candidate completion of Sec.~\ref{sec:uv}, this parity is the remnant of a gauged
$U(1)_N$.
\begin{table}[t!]
\caption{Field content beyond the Standard Model.}\label{tab:fields}
\begin{ruledtabular}\footnotesize
\begin{tabular}{lcccl}
 & $SU(2)_L$ & $U(1)_Y$ & $U(1)_N$ & components\\\hline
$N$      & $\mathbf1$ & $0$         & $+1$ & \\
$N^c$    & $\mathbf1$ & $0$         & $-1$ & \\
$\psi$   & $\mathbf2$ & $-\tfrac12$ & $+1$ & $(\psi^0,\ \psi^-)$\\
$\psi^c$ & $\mathbf2$ & $+\tfrac12$ & $-1$ & $(\psi^{c+},\ \psi^{c0})$
\end{tabular}
\end{ruledtabular}
\end{table}
The doublets form a vector-like pair, as do the singlets. The most general renormalizable
$SU(2)_L\times U(1)_Y\times U(1)_N$ Lagrangian is
\begin{align}
\mathcal L=\mathcal L_{\rm kin}-\Big[\,&\mu\,NN^c+M_D\,\psi\psi^c\nonumber\\
&+y_1\,N^c\psi H-y_2\,N\psi^c\tilde H+{\rm h.c.}\Big],
\label{eq:L}
\end{align}
where $\tilde H=i\sigma_2H^*$ and the $SU(2)$ indices are contracted antisymmetrically; with this
relative sign, both Yukawa terms enter the neutral mass matrix below with the same sign. The only
dimension-three terms that break $U(1)_N$ are
\begin{equation}
\mathcal L_{\not N}=-\tfrac12\big(b_1\,NN+b_2\,N^cN^c\big)+{\rm h.c.}
\label{eq:Lb}
\end{equation}
Hypercharge forbids analogous Majorana masses for the doublets. Thus all $U(1)_N$ violation is
proportional to $b_{1,2}$, making the small splitting technically natural. We impose dark charge
conjugation by taking $y_1=y_2\equiv y$ and $b_1=b_2\equiv b$; the two chiralities then mix equally and the spectrum
is controlled by four real parameters. The imposed $Z_2$ also forbids mixing with Standard-Model
leptons. Radiative stability and the origin of this parity as the remnant of a gauge symmetry are
discussed in Sec.~\ref{sec:uv}.

\subsection{Spectrum}\label{sec:spec}
The charged fields form an unmixed Dirac fermion of mass $M_D$; electroweak loops add the
standard $\Delta M_\pm\simeq355\MeV$ doublet splitting \cite{ThomasWells}.

At $b=0$ the neutral fields form $\PsiN=(N,N^{c\dagger})^T$ and
$\Psip=(\psi^0,\psi^{c0\dagger})^T$. After electroweak symmetry breaking,
\begin{equation}
-\mathcal L_{\rm mass}=\big(\bar\PsiN\ \bar\Psip\big)_L
\!\begin{pmatrix}\mu&m\\ m&M_D\end{pmatrix}\!
\begin{pmatrix}\PsiN\\ \Psip\end{pmatrix}_{\!R}\!+{\rm h.c.},
\label{eq:M2}
\end{equation}
with $m=yv/\sqrt2$, diagonalized by a rotation through
\begin{align}
\tan2\theta&=\frac{2m}{M_D-\mu},\nonumber\\
m_\mp&=\frac{M_D+\mu}{2}\mp\sqrt{\tfrac14(M_D-\mu)^2+m^2}.
\label{eq:theta}
\end{align}
The light state is $\chi=\cos\theta\,\PsiN-\sin\theta\,\Psip$; at small mixing,
$\sin\theta\simeq m/(M_D-\mu)$ and $m_-\simeq\mu-m^2/(M_D-\mu)$.

The Majorana terms split the light Dirac state according to
\begin{equation}
\begin{aligned}
\chi_1&=\frac{i(\chi-\chi^c)}{\sqrt2},&
\chi_2&=\frac{\chi+\chi^c}{\sqrt2},\\
\delta&\equiv m_{\chi_2}-m_{\chi_1}=2b\cos^2\!\theta\simeq2b,
\end{aligned}
\label{eq:delta}
\end{equation}
where $\delta=2b\cos^2\theta$ is the leading result in $b$ and $\delta\simeq2b$ further
neglects the doublet fraction. The splitting is therefore controlled by $b$;
its only dependence on
electroweak symmetry breaking and on the mixing that fixes the LZ rate is the displayed
$\cos^2\theta$ factor. At the reference benchmark of Sec.~\ref{sec:map}, $\chi_1$
is $98\%$ singlet, $\chi_2$ is $300\keV$ heavier, and the remaining neutral and charged states
lie within $0.55\GeV$.

\subsection{Couplings in the mass basis}\label{sec:coup}
The $Z$ couples only to the doublet component. Its vector current becomes off-diagonal when the light
Dirac state is written as two Majorana states,
\begin{equation}
\mathcal L_Z\supset\frac{ig_Z}{2}\st\;Z_\mu\,\bar\chi_1\gamma^\mu\chi_2,
\label{eq:LZ}
\end{equation}
while the charged-current coupling is
\begin{equation}
\mathcal L_W=\frac{g}{\sqrt2}\sin\theta\;W^+_\mu\,\bar\chi\gamma^\mu\psi^-+{\rm h.c.}
\label{eq:LW}
\end{equation}
For equal Yukawa couplings, $y_1=y_2$, both the inelastic-scattering rate and the $WW/ZZ$ annihilation rate scale
as $\sft$. A Yukawa asymmetry would also generate a diagonal axial $Z$ coupling, but the
doublet--singlet degeneracy suppresses the corresponding proton cross section below
$10^{-49}\cmm$ even for an order-one fractional asymmetry. It does not affect the analysis, and
we set $y_1=y_2$.

The diagonal Higgs interaction is
\begin{align}
\mathcal L_h&\supset\frac12g_{h\chi\chi}\,h\bar\chi_1\chi_1,\nonumber\\
g_{h\chi\chi}&=\frac{y}{\sqrt2}\sin2\theta
\simeq\frac{2\st\,(M_D-\mu)}{v}.
\label{eq:ghxx}
\end{align}
For the same coupling choice, $y_1=y_2$, the Higgs interaction is diagonal in $\chi_{1,2}$ and cannot mediate
$\chi_1\leftrightarrow\chi_2$; an off-diagonal coupling requires a spurion that breaks dark charge
conjugation. The Higgs therefore neither drives the inelastic transition nor provides efficient
elastic cooling when the small mass gap required by the relic abundance is imposed.

\subsection{Parameter reduction and benchmark}\label{sec:map}
The mass and recoil spectrum determine $\mu$ and $b$, the relic abundance fixes $M_D-\mu$, and
one LZ event fixes $y$ through the mixing. Our reference benchmark is
\begin{align}
\mu&=730\GeV, & M_D-\mu&=0.19\GeV,\nonumber\\
b&=153.0\keV, & y&=1.6\times10^{-4},\nonumber\\
\delta&=300\keV, & \st&=0.020,\qquad \Omega h^2=0.121.
\label{eq:benchmark}
\end{align}
Thus $m_{\chi_1}\simeq730\GeV$, and the point gives one event in the LZ window. We use it for
intermediate checks and tabulated normalizations; its splitting lies about $3\keV$ below the
central conservative solar ceiling. The point favored by the combined scan is given
separately in Eq.~(\ref{eq:bestfit}).

No additional fields are required at the renormalizable level. Section~\ref{sec:uv} discusses
possible gauge and supersymmetric ultraviolet origins and the remaining unexplained near-degeneracy
$M_D\simeq\mu$; the fractional mass difference is $3\times10^{-4}$ at the reference benchmark.

\section{Constraints and results}\label{sec:calc}
For each $(\mu,\delta)$, the relic abundance fixes $M_D-\mu$ and the LZ normalization fixes
$\st$. The recoil spectrum and the solar neutrino limit then determine whether the point is allowed. Table~\ref{tab:inputs}
lists the inputs; the appendices contain numerical details, sensitivity checks, tests and limitations.
\begin{table}[t]
\caption{Baseline inputs. The nuclear reduced mass $\mu_A$ is evaluated at $m_\chi=730\GeV$.}\label{tab:inputs}
\begin{ruledtabular}\footnotesize
\begin{tabular}{ll}
Input & Value\\\hline
Halo speeds $(v_0,v_{\rm esc})$ & $(238,544)\,{\rm km\,s^{-1}}$\\
Earth speed $v_E(t)$ & $(250+15\cos\omega t)\,{\rm km\,s^{-1}}$\\
Local dark-matter density $\rho_0$ & $0.3\GeV\,{\rm cm^{-3}}$\\
Xenon target & natural isotopes, $\langle A\rangle=131.4$\\
Reduced mass $\mu_A$ ($A=131$) & $104.55\GeV$\\
Helm radius $c$ & $(1.23A^{1/3}-0.6)\,{\rm fm}$\\
Helm parameters $(a,s)$ & $(0.52,0.90)\,{\rm fm}$\\
Exposure & $2.84\,{\rm t\,yr}$\\
Recoil window & $55$--$270\keV$\\
Signal efficiency & published \cite{LZ}; sideband unity
\end{tabular}
\end{ruledtabular}
\end{table}

\subsection{LZ kinematics and normalization}\label{sec:thresh}\label{sec:halo}\label{sec:lzrate}
For endothermic scattering the minimum speed is
\begin{equation}
v_{\min}(E_R)=\frac{1}{\sqrt{2m_AE_R}}\Big(\frac{m_AE_R}{\mu_A}+\delta\Big).
\label{eq:vmin}
\end{equation}
Here $\mu_A$ is the dark-matter--nucleus reduced mass. The minimum of $v_{\min}$ occurs at
\begin{equation}
E_R^\star=\frac{\mu_A\delta}{m_A},\qquad
v_{\min}(E_R^\star)=w_{\min}\equiv\sqrt{\frac{2\delta}{\mu_A}}.
\label{eq:Estar}
\end{equation}
For numerical comparisons we round the reported $248\keV$ recoil to $250\keV$. At $730\GeV$ this gives
$\delta\simeq292\keV$, rather than $290\keV$ at the central recoil energy, with
$w_{\min}=708\,\mathrm{km\,s^{-1}}$ and only $0.45\%$ of the standard halo able to scatter. In the
adopted truncated halo, a $248\keV$ recoil on $A=131$ is inaccessible above $362\keV$ at the
mean Earth speed, and the heaviest isotope at the June maximum extends this to $385\keV$; below
about $243\keV$, the recoil band includes energies below the $55\keV$ threshold.

The differential rate per target nucleus is
\begin{equation}
\frac{d\Gamma}{dE_R}=n_\chi\,\sigma_A\,F^2(E_R)\,\frac{m_A}{2\mu_A^2}\,c^2\,\eta(v_{\min}),
\label{eq:rate}
\end{equation}
with $n_\chi=\rho_0/\mu$, the standard truncated-Maxwellian mean inverse speed $\eta$ averaged
over the Earth's orbital phase, the Helm form factor $F$ \cite{Helm,LewinSmith}, and a sum over
the natural xenon isotopes. The coherent cross section is
\begin{equation}
\sigma_A=\frac{G_F^2\mu_A^2}{2\pi}\big[N-(1-4s_W^2)Z\big]^2\,\sft.
\label{eq:sigA}
\end{equation}
At $730\GeV$ and $\delta=292\keV$, integration over $55$--$270\keV$ with the published
efficiency gives $4.0\times10^3$ events at full coupling, so one event fixes $\st=0.016$ and an
equivalent per-nucleon cross section $5.7\times10^{-43}\cmm$. Curves labeled ``one event''
condition on an expected signal yield of unity; a full signal-plus-background likelihood would
broaden this normalization and the inferred solar limit. A fixed-coupling model must instead
raise the splitting to about $364\keV$. Varying the nuclear form factor moves the inferred $\st$ by
about $10\%$. Uncertainties in the high-velocity tail are larger: a $\pm30\,\mathrm{km\,s^{-1}}$ change in
$v_{\rm esc}$ moves the upper limit on $\delta$ by roughly $\pm10$--$15\keV$. We use the standard
halo \cite{SHM,RAVE}; Appendix~\ref{app:solar} summarizes these sensitivities and the associated
annual modulation.

\subsection{Relic abundance}\label{sec:relic}
The mostly-singlet $\chi_1$ freezes out by coannihilation with the doublet states; electroweak
conversions maintain chemical equilibrium throughout freeze-out. Since $\delta/T_f\sim10^{-5}$,
the Majorana splitting does not affect freeze-out; the relic density is controlled instead by
$\mu$ and $M_D-\mu$ through the thermally populated doublet fraction. In the Griest--Seckel formalism
\cite{GriestSeckel,GondoloGelmini}, with $x\equiv\mu/T$,
$\Delta=(M_D-\mu)/\mu$, $g_1=4$ for the pseudo-Dirac singlet pair and $g_2=8$ for the doublet,
\begin{align}
g_{\rm eff}(x)&=g_1+g_2(1+\Delta)^{3/2}e^{-x\Delta},\label{eq:geff}\\
\langle\sigma_{\rm eff}v\rangle(x)&=\Big[\frac{g_2(1+\Delta)^{3/2}e^{-x\Delta}}{g_{\rm eff}(x)}\Big]^2
\sv_{\rm gauge},\label{eq:seff}
\end{align}
where $\sv_{\rm gauge}=2.4\times10^{-26}\cms(1.1\TeV/M_D)^2$. We choose the coefficient so that
the same relic integral gives $\Omega h^2=0.12$ for a pure doublet at the $1.1\TeV$ Higgsino
thermal mass \cite{MDM,MDMcosmo}. This is an estimate rather than a precision freeze-out
calculation, and the relic-density endpoint below scales with the reference mass. At exact
degeneracy the doublet fraction is $2/3$, giving
$\langle\sigma_{\rm eff}v\rangle=4\sv_{\rm gauge}/9$ and $x_f=25.9$ at $730\GeV$.

Because exact degeneracy maximizes the annihilation rate, $\Omega h^2\le0.12$ \cite{Planck}
implies
\begin{equation}
\mu\le733\GeV
\label{eq:relicbound}
\end{equation}
at zero gap ($733.3\GeV$ in the numerical solution). At each $\mu$, imposing
$\Omega h^2=0.12$ fixes $M_D-\mu$; these points define the relic line. Along it,
$M_D-\mu=2.4\GeV$ at $650\GeV$, $0.93\GeV$ at $700\GeV$, and $0.095\GeV$ at $730\GeV$.
The reference benchmark $(\mu,M_D-\mu)=(730,0.19)\GeV$ is used for normalization and is not
exactly on the relic line: it gives $\Omega h^2=0.121$, one percent above the target and within the
accuracy of the calibrated $\sv_{\rm gauge}$ above. In the scan along the relic line we use
$0.095\GeV$ at $730\GeV$; checks at the reference benchmark retain the $0.19\GeV$ gap.
Section~\ref{sec:degen} discusses the required degeneracy.

\subsection{Solar capture, annihilation and cooling}\label{sec:capture}\label{sec:sv}\label{sec:equil}
The capture rate is the Gould integral \cite{PressSpergel,Gould},
\begin{equation}
C=\int_0^{R_\odot}\!\!4\pi r^2dr\!\int_0^{u_{\max}}\!\!du\;\frac{f(u)}{u}\;w^2
\sum_i n_i(r)\,\sigma_i^{\rm cap}(w,u),
\label{eq:gould}
\end{equation}
with $w^2=u^2+v_{\rm esc}^2(r)$ and $f(u)$ the solar-frame speed distribution normalized to the
local number density, $\int f(u)\,du=n_\chi$. We impose the inelastic threshold, recoil range and capture
condition exactly \cite{Nussinov,Menon}. For the solar potential and composition we use an
analytic profile based on BS05 \cite{BS05}. At $\delta=292\keV$, iron supplies $87\%$ of the rate. For a full
$Z$ coupling, $C=6.8\times10^{23}\,\mathrm{s^{-1}}$; the LZ value
$\st=0.016$ gives $C=1.7\times10^{20}\,\mathrm{s^{-1}}$. Thus $C\propto\sft$ and would exceed
the capture limit inferred from IceCube by a factor four to six if equilibrium held.

Any $\chi_2$ produced by inelastic scattering decays or down-scatters before accumulating, so
solar annihilation proceeds through $\chi_1\chi_1$, without coannihilation. The relevant
$s$-wave channels are $WW$ and $ZZ$, mediated by the nearly degenerate charged and neutral
doublet states. Their amplitudes are $\st$ times the corresponding Higgsino amplitudes; Higgs and
fermionic final states are negligible. Writing $m_1\equiv m_{\chi_1}\simeq\mu$ and
$r_V\equiv m_V^2/m_1^2$, the nonrelativistic result for on-shell gauge bosons and nearly
degenerate propagators is
\begin{equation}
\begin{aligned}
\sv=\sft\Bigg[&\frac{g^4}{128\pi m_1^2}
 \frac{(1-r_W)^{3/2}}{(1-r_W/2)^2}\\[-2pt]
&+\frac{g^4}{256\pi c_W^4m_1^2}
 \frac{(1-r_Z)^{3/2}}{(1-r_Z/2)^2}\Bigg].
\end{aligned}
\label{eq:sv}
\end{equation}
At $m_1=730\GeV$, the coefficient of $\sft$ is $1.80\times10^{-26}\cms$. For
$\delta=292\keV$, the mixing required for one LZ event is $\st=0.016$, which gives
$4.6\times10^{-30}\cms$; the $300\keV$ reference benchmark gives $7.0\times10^{-30}\cms$
(Table~\ref{tab:equil}). Equation~(\ref{eq:sv}) assumes
$m_1>m_Z$; below the gauge-boson thresholds these channels close, so the expression cannot be
continued to $m_1\to0$. We use the $\sft$ rescaling along the nominal solar ceiling in
Fig.~\ref{fig:plane} wherever the Majorana splitting can be treated perturbatively; near the
$733\GeV$ relic-density endpoint we instead use the exact neutral spectrum.

The annihilation rate of the captured population is \cite{GS87,GouldSun,JKG,Catena}
\begin{equation}
\Gann=\tfrac12\,C\tanh^2\!\Big(t_\odot\sqrt{C\sv/V_{\rm eff}}\Big),\;\;
V_{\rm eff}\equiv\frac{\big(\int n\,d^3r\big)^2}{\int n^2\,d^3r},
\label{eq:eq}
\end{equation}
where the equilibration time is $\tau\equiv\sqrt{V_{\rm eff}/(C\sv)}$. A thermal distribution has
$V_{\rm eff}=3.4\times10^{26}\,\mathrm{cm^3}$. We combine the
three-year IceCube conversion between a spin-dependent capture cross section and an annihilation
rate with the ten-year $WW$ cross-section limit and fold in the $WW+ZZ$ mixture of
Eq.~(\ref{eq:sv}) \cite{IceCube3yr,IceCube}. At $0.73\TeV$ this gives
\begin{equation}
\Glim(0.73\TeV)=1.8\times10^{19}\ {\rm s^{-1}},
\label{eq:glim}
\end{equation}
with a recast range $(1.4$--$2.3)\times10^{19}\,\mathrm{s^{-1}}$.
Appendix~\ref{app:solar} gives the details. IceCube likelihoods
for inelastic dark matter, including the dependence on whether the captured population thermalizes,
exist for lighter masses and splittings up to a few hundred keV \cite{Catena,CRRS}. A dedicated
event-level application at $0.73\TeV$ with the ten-year sample would refine
Eq.~(\ref{eq:glim}). At fixed electroweak strength, solar capture can remain efficient up to
splittings near $550\keV$ at TeV masses
\cite{DiMauroShaikh2026,NguyenLindenHooper2026,BoseEtAl2026}; our lower
rate ceiling follows because the one-event normalization suppresses both capture and annihilation.

At gauge strength the argument of the $\tanh$ is about 850 and the Sun is in equilibrium. Here
both $C$ and $\sv$ scale as $\sft$, so that argument becomes $850\sft\simeq0.3$ at the
reference benchmark. In the non-equilibrium limit,
\begin{equation}
\Gann\ \longrightarrow\ \tfrac12\,t_\odot^2\,\frac{C^2\,\sv}{V_{\rm eff}}\ \propto\ \sin^{12}\theta,
\label{eq:quad}
\end{equation}
where $\Gann\propto(\sft)^3=\sin^{12}\theta$, compared with $\sft$ for LZ. The reference
benchmark has an equilibration time of about $14$\,Gyr, rather than $5$\,Myr at gauge strength.
Solving $\Gann=\Glim$ with the thermal $V_{\rm eff}$ gives the conservative solar ceiling,
$\delta=302$--$305\keV$ at $730\GeV$ (Table~\ref{tab:equil}).

An inelastically captured solar population need not thermalize \cite{BCH16,BCH18}. In our model,
inelastic scatters stop near the MeV scale, after which only the doubly suppressed Higgs coupling
in Eq.~(\ref{eq:ghxx}) can cool the orbit.
In the isotropic cooling ansatz, $V_{\rm eff}$ is $(2.2$--$3.9)\times10^3$ times the thermal
value at the reference benchmark and the annihilation rate is reduced by
$(2.7$--$4.9)\times10^{-4}$. The Higgs interaction is the suppressed coupling in
Eq.~(\ref{eq:ghxx}), rather than an unsuppressed diagonal scalar interaction that can thermalize
other pseudo-Dirac candidates \cite{DiMauroShaikh2026}. Appendix~\ref{app:solar} gives the cascade
and cooling calculation.

After imposing the relic abundance at each mass, the nominal solar ceiling reaches $331\keV$ at
$730\GeV$ and $343\keV$ at the $733\GeV$ relic-density endpoint. A transport calculation following
individual scatterings changes these ceiling values by at most $1.7\keV$
(Appendix~\ref{app:transport}). Figure~\ref{fig:plane} uses the tree-level elastic rate;
Appendix~\ref{app:loop} gives the loop correction and its sign uncertainty. The conservative
ceiling is independent of elastic cooling.

\subsection{LZ spectrum and sideband}\label{sec:shape}
\begin{figure}[tbp]
\centering
\includegraphics[width=\columnwidth]{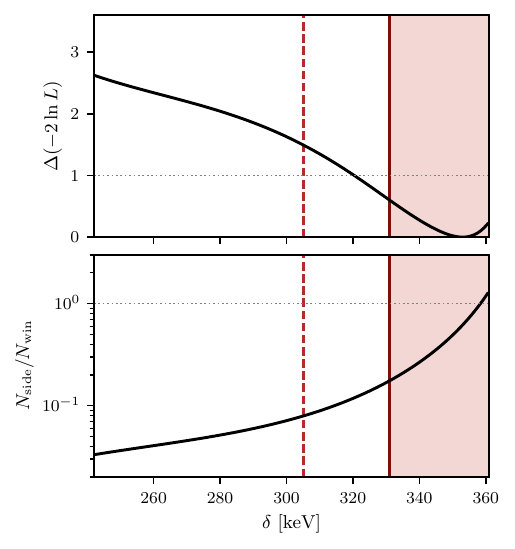}
\caption{Likelihood penalty from the LZ recoil shape (top) and predicted sideband events per event
in the LZ window (bottom) at $730\GeV$.
The solid and dashed vertical lines mark the nominal and conservative solar ceilings, respectively.}
\label{fig:lzshape}
\end{figure}

\begin{figure}[tbp]
\centering
\includegraphics[width=\columnwidth]{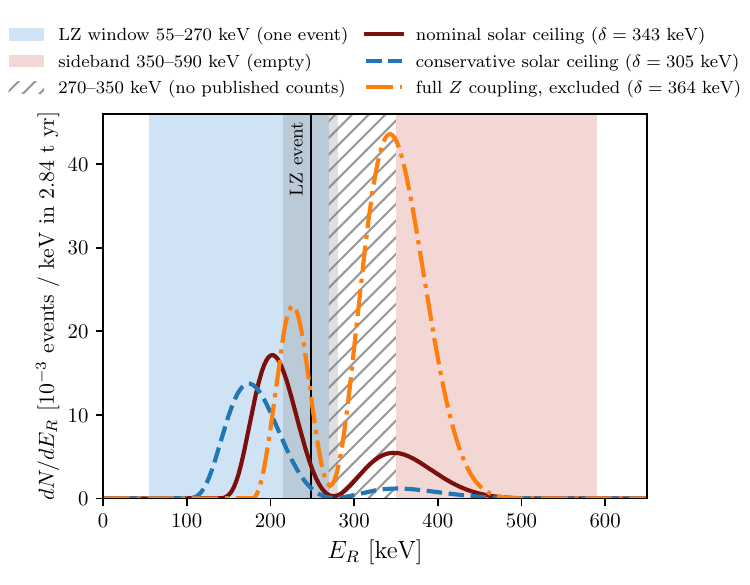}
\caption{Xenon recoil spectra before the detector efficiency, normalized to one detected event
in the LZ window. Solid: the nominal solar ceiling $(733\GeV,343\keV)$. Dashed: the conservative
ceiling $(730\GeV,305\keV)$. Dash-dotted: the solution with the full $Z$ coupling
$(730\GeV,364\keV)$. With unit
acceptance in the sideband they predict respectively $0.31$, $0.08$, and $1.9$ sideband events.}
\label{fig:spectra}
\end{figure}

The event energy and the empty $350$--$590\keV$ sideband constrain $\delta$ independently of
the normalization. Increasing $\delta$ improves the recoil-energy fit up to about $350\keV$. At
larger splittings the empty sideband adds a likelihood penalty, although the present penalty remains
mild near $350\keV$ (Fig.~\ref{fig:lzshape}) \cite{LZ,RSSX,DentNewstead}. After normalizing each spectrum to one event in the LZ
window, $N_{\rm side}/N_{\rm win}$ is the predicted number of sideband events per event in the
LZ window, and
$\Delta_{\rm shape}\equiv\Delta(-2\ln L)$ is the profile-likelihood penalty relative to its
unconstrained minimum. No counts are published between 270 and $350\keV$, and we do not use that
interval.
Figure~\ref{fig:spectra} compares the corresponding spectra at the two solar ceilings and the point
normalized to one event with the full $Z$ coupling.
After smearing with the quoted resolution, the median energy of recoils reconstructed inside the
$55$--$270\keV$ window moves from $168\keV$ at $\delta=270\keV$ to $206\keV$ at $345\keV$
(Table~\ref{tab:side}). Raising $\delta$ therefore
improves the shape, but normalizing the rate to one event forces $\sft$ upward as the halo fraction
falls. Both capture and equilibration then become more efficient, producing the steep solar limit shown in
Fig.~\ref{fig:plane}.

\subsection{Combined parameter space}\label{sec:corner}
Figure~\ref{fig:plane} combines the relic-density endpoint, the LZ kinematic and spectral constraints,
the sideband constraint, and the solar limits. The recoil likelihood favors larger $\delta$ until
the nominal solar limit is saturated, yielding
\begin{equation}
\begin{split}
&\mu\simeq730\text{--}733\GeV,\quad \delta\simeq331\text{--}343\keV,\\
&\st\simeq0.065\text{--}0.127,\quad M_D-\mu\simeq10\text{--}95\MeV,
\end{split}
\label{eq:bestfit}
\end{equation}
with $\Gann=\Glim$ and $\Delta_{\rm shape}=0.2$--$0.6$. Points below the nominal curve remain
allowed by the constraints considered here. The conservative limit for a thermalized population
rises from about $291\keV$ at $500\GeV$ to $302$--$305\keV$ near the relic-density endpoint;
inefficient cooling raises the nominal limit there by $28$--$39\keV$.
Near the relic-density endpoint, viability requires both the small Higgs coupling associated with
the mass gap fixed by the relic abundance and the suppression of both capture and annihilation by
the same mixing. Near degeneracy, adding the
$\chi_1$--doublet channels does not move the relic-density endpoint because the coannihilation sum is a
trace over the degenerate states. Appendix~\ref{app:tests} gives the resulting predictions.

Because the preferred point saturates the solar limit, its shape penalty is a goodness-of-fit
statement, not a revision of LZ's $3.4\sigma$ local or $2.6\sigma$ global significance.\label{sec:signif}
At the nominal ceiling, $\Delta_{\rm shape}=0.2$--$0.6$; at the conservative ceiling it is $1.6$
(Fig.~\ref{fig:lzshape}). Thus the recoil data favor larger $\delta$, while the solar result sets its
upper limit.

\section{Possible ultraviolet completions}\label{sec:uv}
The low-energy theory is renormalizable. Its ultraviolet origin raises four questions: the
origin of the stabilizing parity and Majorana splitting, the $O(10^{-4})$ Yukawa coupling,
whether dark charge conjugation is a symmetry limit, and whether a supersymmetric embedding
exists. Table~\ref{tab:uv} summarizes several possibilities; none answers all four.

\begin{table*}[t]
\caption{Candidate ultraviolet completions. ``Endpoint'' is the largest mass in TeV consistent with
the calibrated relic-density estimate; the minimal model reaches $0.73$
(Sec.~\ref{sec:relic}). ``Fixed'' means $y=g'/\sqrt2$. The last column lists the near-degenerate
masses that lack a common origin, as discussed in Sec.~\ref{sec:degen}.}
\label{tab:uv}
\begin{ruledtabular}
\begin{tabular}{lccccl}
 & $\delta$ protected & $y\simeq10^{-4}$ & $y_1=y_2$ & endpoint [TeV] & mass relation\\
\hline
gauged $U(1)_N$            & yes   & no    & imposed & 0.73 & $M_D\simeq\mu$\\
doublet $\Phi$ with induced vev & yes   & yes   & imposed & $\geq0.73$ & $M_D\simeq\mu$\\
$R$-symmetric bino $+\tilde\ell_R$ & yes & fixed & $N{=}2$ & 0.58 & $\mu_d\simeq m_D$\\
$R$-symmetric bino $+\tilde W_D$ & yes & fixed & $N{=}2$ & 0.9 & $\mu_d\simeq m_D$ and $m_{\tilde W}\simeq m_D$\\
$R$-symmetric singlet pair & yes   & free  & no      & 0.73 & $\mu_d\simeq\mu_N$\\
Majorana gauginos & $M\!\gtrsim\!500\TeV$ & no & no & --- & split spectrum\\
\end{tabular}
\end{ruledtabular}
\end{table*}

\emph{Gauge and scalar origins.} Gauging $U(1)_N$ and breaking it with a charge-two singlet
$S$ leaves a stabilizing $Z_2$ remnant \cite{KraussWilczek}. The terms
$y_SS^*NN+y'_SSN^cN^c$ generate $b_{1,2}$, whose smallness is technically natural because Dirac
number is restored as $y_S,y'_S\to0$. Dark charge conjugation at the matching scale
can impose $y_1=y_2$ and $b_1=b_2$; radiative violation is negligible near the degeneracy required
by the relic abundance.

The small Yukawa can arise from an induced vacuum expectation value
\cite{Ma2001,DavidsonLogan} or from an exponentially suppressed wavefunction overlap in an extra
dimension \cite{AHS,MS99,GN,GP,KT00,KT01,DKYY}. If the dark fermions couple with order-one strength to a second
doublet $\Phi$, a soft term $m_{12}^2H^\dagger\Phi$ gives
$v_\Phi\simeq(m_{12}^2/M_\Phi^2)v$ and $y_{\rm eff}=yv_\Phi/v$. The required
$y_{\rm eff}=1.6\times10^{-4}$ corresponds to $v_\Phi\simeq40\MeV$, or
$m_{12}\simeq12\GeV$ for $M_\Phi=1\TeV$. The induced vacuum expectation value leaves the
low-energy $Z$ coupling unchanged. For the $1\TeV$ illustration, however, additional annihilation
and coannihilation channels involving $\Phi$ are open at freeze-out; the relic and solar results
apply only if these channels are subdominant, so $M_\Phi>m_\chi$ alone is insufficient.

\emph{Supersymmetric possibilities.} A Majorana mass $m_M$ for the doublet partner induces
$\delta\simeq\st m_M$. With Majorana gauginos, $m_M\simeq m_Z^2/M_{1,2}$ and hence
\begin{equation}
 \delta_{\rm induced}\simeq\st\,\frac{m_Z^2}{M}\simeq0.2\GeV\Big(\frac{1\TeV}{M}\Big)
\label{eq:majfeed}
\end{equation}
at the mixing required for one LZ event. Obtaining $300\keV$ requires $M\gtrsim500\TeV$, favoring a split
spectrum or Dirac gauginos \cite{DiracNMSSM,FNW,KPW}. In the bino--wino regime we examined,
Higgsino exchange fixes the off-diagonal mass and $Z$ coupling together, so the transition
strength cannot be suppressed independently.

An $R$-symmetric Dirac bino--adjoint pair and vector-like Higgsino pair have, in the basis
$(\tilde B,\tilde R_d^0)\times(\tilde S,\tilde H_d^0)$, the block
\begin{equation}
 {\cal M}_d=
 \begin{pmatrix}
  m_D & g'v_d/2\\
  \lambda_dv_d/\sqrt2 & \mu_d
 \end{pmatrix},
 \label{eq:susyblock}
\end{equation}
where $v_d=v\cos\beta$. At the $N=2$ boundary $\lambda_d=g'/\sqrt2$ \cite{BGS}, the block is
Eq.~(\ref{eq:M2}) with $(\mu,M_D,v,y)\to(m_D,\mu_d,v_d,g'/\sqrt2)$; small $R$-breaking
Majorana masses supply the splitting \cite{Hsieh}. With the coupling fixed at gauge strength,
requiring one event gives $\mu_d-m_D\simeq6$--$11\GeV$, $\tan\beta\simeq38$--$58$, and
$m_D\simeq0.29$--$0.53\TeV$. For this candidate, the thermalized solar limit of
$267$--$292\keV$ lies below the $305$--$355\keV$ range preferred by the recoil spectrum. Slepton
and Dirac-wino coannihilation can
extend the allowed mass range, as shown in Table~\ref{tab:uv}; Appendix~\ref{app:susycoann} gives
the calculation.

The Dirac neutralino matrix admits two physical sign choices. The branch in
Eq.~(\ref{eq:susyblock}) gives level repulsion and
$\st\simeq g'^2v_d^2/[4(\mu_d-m_D)^2]$; the other replaces the small mass difference by a sum and
gives $\st\sim10^{-6}$ in the quoted region. In the conventions of Ref.~\cite{DKKS}, the required
$N=2$ branch has $\operatorname{sign}(m_D)=-\operatorname{sign}(\mu_d)$; this relative sign is an
additional discrete choice.

Diagonalizing the full $4\times4$ neutralino matrix in Eq.~(2.25) of Ref.~\cite{DKKS} confirms the
two-state reduction. The bino row and singlet column have no direct wino or triplino entry, so a
nearby Dirac wino enters neither mixing angle at leading order. Bringing it within $5$--$10\GeV$
of the bino changes $\st$ by less than one percent, while the state remains $95\%$ bino and the
diagonal axial coupling vanishes at the $N=2$ point.

Taking the dark matter to be singlet-like rather than bino-like removes the fixed relation between
the Yukawa and gauge couplings. Singlet superfields
$N,N^c$ with $R$ charges $(0,2)$ admit
$W\supset\mu_NNN^c+\lambda NR_dH_d+\lambda'N^cH_uH_d$. Their neutral fermion mass block reproduces
Eq.~(\ref{eq:M2}) under $y_1v\to\lambda'v_u$ and $y_2v\to\lambda v_d$, apart from a
$\lambda'v_d/\sqrt2$ mixing outside the block. This candidate retains the full low-energy
parameter space, but neither enforces $\lambda'\tan\beta=\lambda$ nor explains
$\lambda'\sim10^{-4}$.

The supersymmetric entries are candidate embeddings. Establishing them requires a consistent
treatment of the charged sector, collider limits, loop-induced scattering, and the Higgs mass in
the same spectrum, together with precision relic and solar calculations. A weakly coupled $Z'$
remains a non-supersymmetric alternative \cite{DiMauro,LeeRandall2026}.

All candidates still require an unexplained near-degeneracy.\label{sec:degen} The minimal model requires
$M_D-\mu\lesssim0.2\GeV$ at the reference benchmark and $10$--$95\MeV$ near the nominal
ceiling; the $R$-symmetric bino relaxes this to about $6\GeV$, but instead requires a Dirac gaugino
mass close to a superpotential mass with the opposite sign, and also a nearby Dirac wino to reach
the $0.73\TeV$ region. No construction in
Table~\ref{tab:uv} supplies a common origin for these masses.

\section{Conclusions and outlook}\label{sec:fate}
A Majorana singlet mixed with a vector-like doublet can reproduce the LZ event while satisfying
the relic and solar constraints considered here. The mixing suppresses both scattering and
annihilation, preventing solar equilibrium, while the small mass gap required for coannihilation also weakens
elastic cooling. A thermalized population gives the robust upper limit $\delta=302$--$305\keV$;
within the cooling ansatz, the nominal upper limit rises to $331$--$343\keV$ near the relic-density
endpoint, which is also favored by the recoil spectrum.

The comparison with other inelastic interpretations gives a broader model-building lesson. A viable
mediator must avoid a diagonal elastic interaction, obtain the required suppression from the
transition coupling rather than by increasing $\delta$, and suppress annihilation by the same
factor unless the captured population remains spatially extended or annihilation is velocity
suppressed. A Majorana vector
current satisfies the first condition without tuning. Within Standard-Model mediators, the three
requirements point to an off-diagonal $Z$ interaction generated by singlet--doublet mixing. A
weakly coupled $Z'$ provides an alternative transition interaction \cite{DiMauro,LeeRandall2026},
but the same requirements must then be checked in the complete model.

The model therefore predicts correlated solar, xenon, and target-dependent signals,
with no observable tree-level elastic scattering in the preferred $730$--$733\GeV$ region.
Benchmark rates and collider signatures are
given in Appendix~\ref{app:tests}; the assumptions entering the nominal solar limit are
collected in Appendix~\ref{app:solar}. The model-building
question left open is the origin of the near equality $M_D\simeq\mu$ identified in
Sec.~\ref{sec:degen}.

\begin{acknowledgments}
The authors thank Jae Hyeok Chang and Lisa Randall for useful discussions and feedback.
SJL and TY are funded by the Samsung Science and Technology Foundation under Project Number
SSTF-BA2201-06. TY is also supported by the Gravity, Spacetime, and Particle Physics (GRASP)
Initiative from Harvard University.
\end{acknowledgments}

\appendix
\section{Solar, likelihood and systematic checks}\label{app:solar}
\subsection{Nonthermal solar population}\label{app:transport}
Inelastic scattering reduces a captured particle's kinetic energy to about $1\MeV$, still far
above the solar temperature. We evolve an isotropic population from that energy under
orbit-averaged Higgs scattering. At the reference benchmark the resulting
$V_{\rm eff}$ is $(2.2$--$3.9)\times10^3$ times the thermal value, depending on the starting
distribution, and $\Gann$ is $(2.7$--$4.9)\times10^{-4}$ of the thermalized result.
Simulations at smaller $\delta$ and $m_\chi$ with no elastic channel find the same qualitative
effect \cite{BCH18}. Equation~(\ref{eq:eq}) is exact for a fixed spatial profile. Allowing the
capture-age mixture to evolve and accounting for preferential depletion of its denser, older
components changes $\Gann$ by $0.1\%$ or less at the three checked points and the nominal ceiling
by less than $0.002\keV$.

To test the isotropic treatment, a second transport calculation samples the capture kinematics and
tracks the energy and angular momentum through individual inelastic and elastic scatters. The
excited state produced in each inelastic scatter either decays before its next collision or, with
the tree-level lifetime of Appendix~\ref{app:tests}, down-scatters exothermically first when that
is faster; the two treatments agree within the ranges quoted here. At three checked points
between $730$ and $733\GeV$, the transported population has an effective volume $V_{\rm eff}$ a factor
$1.8$--$3.5$ larger than in the isotropic treatment. Because $\Gann$ rises steeply with $\delta$, the nominal ceiling moves by only
$1.4$--$1.7\keV$. At $725\GeV$ about $15\%$ of the population reaches the thermal regime, where
this calculation ceases to apply. Below that mass, thermal target motion and detailed balance are
needed before a correction can be assigned, so the main text uses the original curve from the
isotropic treatment.
The curve for the nominal solar limit in Fig.~\ref{fig:plane} stops at $733\GeV$, where the two-state
approximation for the transition coupling is accurate to $3\%$ and the exact spectrum changes the
limit by $0.1\keV$; beyond this mass the nearly degenerate neutral states reorder.

\subsection{Systematic uncertainties and elastic scattering at one loop}\label{app:loop}
The largest numerical sensitivity is the high-velocity halo tail. At $\delta=292\keV$, lowering or
raising $v_{\rm esc}$ by $30\,{\rm km\,s^{-1}}$ changes $\Gann$ by factors $8.6$ and $0.30$,
respectively; varying
$v_0$ from $238$ to 220 or $250\,{\rm km\,s^{-1}}$ changes it by factors $5.8$ and $0.38$.
The same tail produces June-to-December rate ratios of
$1.8$, $2.6$, $5.9$ and $\gg10$ at $\delta=250$, 292, 320 and $350\keV$, respectively.

Full one-loop spin-independent calculations exist for singlet--doublet models where the tree-level
Higgs coupling vanishes \cite{HLMW}. A leading-order scale estimate obtained by rescaling
pure-doublet coefficients \cite{HisanoLoop,HillSolonI,HillSolonII,ChenHill,PR} gives
$\sigma_{\rm SI}^{\rm loop}\simeq\sft\sigma_H$, with
$\sigma_H=(0.4$--$1)\times10^{-49}\cmm$. At the reference gap,
$M_D-\mu=0.19\GeV$, this is $(2$--$4)\times10^{-53}\cmm$, below the tree value
$5.6\times10^{-52}\cmm$. Along the relic line the two amplitudes become comparable at
$M_D-\mu\simeq28$--$46\MeV$; the loop estimate dominates below about $13\MeV$ and reaches
$(0.6$--$1.6)\times10^{-51}\cmm$ at $733\GeV$, where the exact tree value is
$7.7\times10^{-53}\cmm$.

We also evaluated the electroweak boxes and Higgs penguins in the mass basis, including the leading
gluon matching. For the central inputs the loop and tree amplitudes have the same sign, lowering the
nominal ceiling by about $3$--$8\keV$ over $730$--$733\GeV$. Hadronic and matching uncertainties do
not fix the relative sign; with the opposite sign, a cancellation lets the ceiling approach the no-cooling
value near $343\keV$ before it falls again. We therefore quote the tree-level result. The
conservative thermalized ceiling is independent of elastic cooling.

A $\pm30\,{\rm km\,s^{-1}}$ variation of $v_{\rm esc}$ moves the solar ceiling by roughly
$\pm10$--$15\keV$. The remaining quoted systematics are smaller: a $15\%$ uncertainty in the
solar iron abundance changes $C$ by about $13\%$, and a $30\%$ change in the recast IceCube limit
moves the conservative ceiling by about $2\keV$ and the nominal one by less than $1\keV$. A full
experimental treatment still requires the unpublished LZ response above the analysis window and
an IceCube likelihood recast.

\subsection{Solar limit and recoil likelihood}
For a thermalized population, Table~\ref{tab:equil} evaluates Eq.~(\ref{eq:eq}) at
$\mu=730\GeV$, fixing the mixing to one LZ event at each splitting. At full electroweak strength
$t_\odot/\tau\simeq850$ and equilibrium is automatic; the reduced mixing instead leaves the
parameter region of interest out of equilibrium.
\begin{table}[t]
\caption{Thermalized solar population at $\mu=730\GeV$. The mixing is fixed by one LZ event and
$\Glim=1.8\times10^{19}\,{\rm s^{-1}}$, with a recast range
$(1.4$--$2.3)\times10^{19}\,{\rm s^{-1}}$.}\label{tab:equil}
\begin{ruledtabular}\scriptsize
\begin{tabular}{lrrrr}
$\delta$ [keV] & 270 & 292 & 300 & 320\\\hline
$\st$ & 0.0098 & 0.0159 & 0.0197 & 0.0389\\
$\sv$ [$10^{-30}\cms$] & 1.7 & 4.6 & 7.0 & 27\\
$C$ [$10^{20}\,{\rm s^{-1}}$] & 0.85 & 1.73 & 2.39 & 7.06\\
$t_\odot/\tau$ & 0.093 & 0.21 & 0.31 & 1.06\\
$\tanh^2(t_\odot/\tau)$ & 0.0085 & 0.045 & 0.092 & 0.62\\
$\Gann/\Glim$ & 0.016--0.026 & 0.17--0.28 & 0.48--0.79 & 9.5--15.6
\end{tabular}
\end{ruledtabular}
\end{table}
The limit in Eq.~(\ref{eq:glim}) uses the three-year IceCube conversion from a spin-dependent
cross section to an annihilation rate and the ten-year $WW$ cross-section limit at each mass
\cite{IceCube3yr,IceCube}. Folding the $WW+ZZ$ spectrum through the ten-year effective area gives
the central value and recast range quoted above; a full event-level likelihood is not public.

For the recoil likelihood we convolve the spectrum with a $23\keV$ resolution, profile the
$23\keV$ energy-scale nuisance, and include Poisson terms for the empty low-energy region and the
published $350$--$590\keV$ sideband. Table~\ref{tab:side} gives the quantities used in
Fig.~\ref{fig:lzshape}. Reasonable Helm and Fermi form-factor choices change
$\Delta(-2\ln L)$ by less than $0.1$.
\begin{table}[t]
\caption{Predicted sideband yield and likelihood from the recoil spectrum at $m_\chi=730\GeV$, after normalizing
to one event in the LZ window.}\label{tab:side}
\begin{ruledtabular}\scriptsize
\begin{tabular}{lrrrrr}
$\delta$ [keV] & 270 & 292 & 320 & 330 & 345\\\hline
$N_{\rm side}/N_{\rm win}$ & 0.045 & 0.062 & 0.12 & 0.17 & 0.35\\
median $E_R$ in $55$--$270\keV$ & 168 & 174 & 187 & 194 & 206\\
$\Delta_{\rm shape}$ & 2.20 & 1.81 & 1.01 & 0.64 & 0.13
\end{tabular}
\end{ruledtabular}
\end{table}

\section{Experimental signatures}\label{app:tests}
\subsection{Solar and recoil tests}
After imposing the relic abundance and one LZ event, the allowed parameter region lies below the
nominal solar limit in Fig.~\ref{fig:plane}. At that limit the solar neutrino rate reaches the
IceCube limit; at the reference benchmark, with the nonthermal population of
Sec.~\ref{sec:equil}, it is $(1$--$4)\times10^{-4}$ of the limit, spanning the IceCube recast band
and the two starting distributions, because it falls as $\sin^{12}\theta$. A stronger IceCube
limit would first lower the best-fitting allowed splitting.
Under the thermalized assumption, an improvement by a factor of a few hundred would exclude the
remaining parameter space; excluding the reference benchmark with a nonthermal population would
take a factor of a few thousand.

The next xenon recoils test the splitting directly. Their median energy inside the LZ window,
after smearing, is $204\keV$ at $\delta=343\keV$ and $177\keV$ at the reference benchmark; the
corresponding June-to-December ratios are 56 and 3. Strong seasonal modulation of inelastic
interpretations of the LZ event has also been emphasized in Ref.~\cite{McCabe}. Several further
events above $240\keV$ would make the recoil spectrum favor splittings above the solar limit,
while events below $150\keV$ would favor smaller splittings. A low-energy elastic
population correlated with the high-energy recoils would exclude the preferred
$730$--$733\GeV$ region.

Target dependence is an independent test. Tungsten gives 26 events per tonne-year at the
nominal ceiling and 3 at the reference benchmark, about $46$ and $7$ times the xenon recoil rate
before detector efficiency, while germanium, argon and silicon cannot open the inelastic channel
anywhere in the halo.
A dedicated CaWO$_4$ exposure at high recoil energies could test this hierarchy \cite{CRESST}; any
confirmed signal at high recoil energy in germanium or argon would rule it out. These target ratios do not depend on the mixing,
relic history or solar calculation.

\subsection{Collider and complementary searches}
The nearly degenerate doublet states provide the model's characteristic collider signature. Near
the nominal solar limit, $M_D-\mu=10$--$95\MeV$, with four neutral states and one charged state
within about $460\MeV$. At the reference
benchmark, the charged and heavy neutral states have decay lengths of about $0.6$\,cm and $1$\,m,
respectively. The heavy neutral state decays both by photon emission \cite{KrallReece} and through
a neutral pion to the light states; the charged-state decay length is set by the radiative
charged--neutral splitting and hardly changes. The heavy neutral decay length grows to kilometers
as $M_D-\mu$ falls below the pion mass near the nominal solar limit, and both states evade present
compressed-spectrum searches, whose tracklet reach is limited by the short charged-state lifetime
\cite{GKMY}. Probing a $730\GeV$
doublet likely requires a future hadron or muon collider
\cite{SaitoDT,HanMuC}.

Other searches are much less sensitive. Higgs exchange gives
$\sigma_{\rm SI}=5.6\times10^{-52}\cmm$ at the reference benchmark and
$7.7\times10^{-53}\cmm$ at $733\GeV$,
far below the neutrino fog \cite{NeutrinoFog}; the
loop cross section is $\sft$ times the pure-doublet value \cite{HLMW}, reaching up to
$1.6\times10^{-51}\cmm$ near the $733\GeV$ relic-density endpoint
(Appendix~\ref{app:loop}). The present-day
annihilation rate is also far below the thermal value and beyond CTA sensitivity \cite{CTA}. At the reference benchmark the tree-level
$\chi_2\to\chi_1\nu\bar\nu$ lifetime is about $100$ years; an allowed transition dipole could
shorten it, but either channel removes the excited population before recombination.

Along the relic line, the one-event normalization together with the LZ spin-independent limit
excludes masses below about $460\GeV$ \cite{LZWIMP}. This lower bound is independent of the solar
calculation and lies well below the preferred high-mass region.

The characteristic experimental pattern is therefore a solar signal near the upper limit on
$\delta$, additional xenon recoils at somewhat lower energies, a much larger tungsten rate, no
germanium or argon signal, and no observable tree-level elastic scattering in the preferred
$730$--$733\GeV$ region.

\section{Supersymmetric coannihilation estimate}\label{app:susycoann}
To obtain the $R$-symmetric relic-density endpoints in Table~\ref{tab:uv}, we extend
Eqs.~(\ref{eq:geff})--(\ref{eq:seff}) to a Dirac bino, Higgsinos, right sleptons and a possible
Dirac wino. We retain $s$-wave channels at threshold and neglect mixing- and
Yukawa-suppressed terms.
We include Dirac-bino annihilation through sleptons \cite{BHK}, bino--slepton coannihilation
\cite{EFOS}, slepton-pair gauge annihilation and Dirac-wino annihilation. For
$\mu_d-m_D=6\GeV$, degenerate sleptons lower the relic-density endpoint to $0.32\TeV$, whereas
$m_{\tilde\ell}=1.15$--$1.2\,m_D$ raises it to $0.58\TeV$, above its value without sleptons.
A Dirac wino $4$--$18\GeV$ above the bino allows $m_D=0.73\TeV$, and the degenerate limit
reaches $0.9\TeV$. These endpoint values use the same approximate coannihilation treatment as
Sec.~\ref{sec:relic}; a complete mixed neutralino calculation is left for future work.

\section{Comparison with other interpretations}\label{app:readings}
Table~\ref{tab:census} applies the same LZ normalization and solar test to representative
inelastic interpretations. For fixed electroweak couplings we use the thermalized population
appropriate when an elastic loop channel is present. These are our calculations under common
assumptions, not exclusions claimed by the cited authors. The last row is evaluated at points that
saturate this model's nominal solar limit; points with smaller $\delta$ have smaller rates.
\begin{table}[t]
\caption{Solar comparison after normalizing to one event in the LZ window. Masses and splittings are
in GeV and keV. For the triplet--doublet model \cite{HCMDM}, we use its $2.9\TeV$ thermal relic
with a transition current of Higgsino strength; its published normalization is four times smaller.
The dark-photon row uses the pure-$WW$ limit; its recast-band range is $1.7$--$100$. The final row
corresponds to this model's nominal solar limit.}\label{tab:census}
\begin{ruledtabular}\scriptsize
\begin{tabular}{@{}lrrr@{}}
interpretation & $m_\chi$ & $\delta$ & $\Gann/\Glim$\\\hline
Higgsino \cite{Higgsino,WZZ,DiMauro,FanReece} & 1080 & 375 & $4.2\times10^3$\\
$Z_2$ doublet \cite{Nomura} & 440--600 & 341--357 & $(0.75$--$0.88)\times10^4$\\
PQ doublet \cite{Vis} & 287--509 & 304--344 & $(0.4$--$2.0)\times10^3$\\
triplet--doublet model \cite{HCMDM} & 2900 & 386 & $6.2\times10^2$\\
dark photon \cite{DP} & 420 & 250--313 & $2.2$--$78$\\\hline
this work & 730--733 & 331--343 & 1
\end{tabular}
\end{ruledtabular}
\end{table}

Fixed-coupling Higgsino and doublet models need a larger splitting to normalize the LZ rate, but
solar acceleration removes this halo-tail suppression; a subcomponent rescales LZ and capture
together \cite{PR,Nomura,Vis}. An inert-doublet fit lies in the same region \cite{WangXiao2026}.
In the triplet--doublet model, loop-induced elastic scattering thermalizes the captured population
\cite{HCMDM}. For its $2.9\TeV$ thermal relic, the splitting required by LZ exceeds the solar limit
and shifts xenon recoils toward the sideband \cite{RSSX}. The analysis of
Ref.~\cite{HCMDM} does not include our singlet--doublet model.

Di Mauro's pseudo-Dirac model is the closest precedent: an effective weak $Z'$ suppresses the
transition \cite{DiMauro}, and the captured population can remain out of equilibrium
\cite{DiMauroShaikh2026}. A gauged baryon-flavor model has the same suppression, with prompt solar
neutrinos arising only through a loop-induced $Z'W^+W^-$ coupling \cite{HMLee2026}. In a
dark-photon model, $WW$ annihilation brings the captured population into capture--annihilation
equilibrium, and the annihilation rate exceeds the IceCube limit throughout the range in
Table~\ref{tab:census}. The construction separates diagonal and transition scalars; a discrete
symmetry enforcing this separation is left for future work \cite{DP}. An approximate left--right
interchange symmetry separates the diagonal and transition scalar couplings in another light-scalar
model \cite{DasEtAl2026}.

Other proposals invoke model-independent analyses, halo uncertainties, boosted or exothermic dark matter,
absorption or backgrounds, an elastic pseudoscalar portal, a transition magnetic dipole,
nuclear-response interference, or a companion signal from xenon de-excitation
\cite{SYY,FanReece,McCabe,LouLu,AtmNu,Unwin,deLima,GuLiTangXu,DentNewstead,KhanEtAl2026,He2026,FanEtAl2026,ChattarajEtAl2026,AlhazmiEtAl2026}.
An elastic spin-dependent interpretation with the same field content has also been proposed
\cite{ElahiSchwaller2026}. Its low-energy spectrum differs from ours: $y_1=y_2$ removes the
diagonal $Z$ coupling at tree level while leaving a mixing-suppressed transition.
Higgsino and singlino embeddings have also been discussed \cite{DuWang,Yin,NagataShirai,Singlino}.
In the general NMSSM, singlino mixing can cancel the gaugino contribution to the splitting and
reduce the $Z$-mediated inelastic rate by $10$--$20\%$, much less than the suppression required here
\cite{BisalCaoLi2026}.
The singlet--doublet model is standard \cite{MS06,DEramo,EFHPP,CKPT,CMT,BMMT,SD3,SD1,KMSS21,GKSS22,SD2,SDrev};
here its splitting is chosen to make the transition visible rather than merely to evade direct detection.

%% file: refs.bib
@control{apsrev42Control,
    author = "08",
    editor = "1",
    pages  = "0",
    title  = "1",
    year   = "1"
}

@article{LZ,
    author = "Akerib, D. S. and others",
    collaboration = "LZ",
    title = "{Search for dark matter particle interactions in an extended nuclear recoil energy window with the LUX-ZEPLIN (LZ) experiment}",
    eprint = "2609.02823",
    archivePrefix = "arXiv",
    primaryClass = "hep-ex",
    month = "9",
    year = "2026",
    journal = ""
}

@article{PR,
    author = "Pospelov, Maxim and Ramani, Harikrishnan",
    title = "{Strong Constraints on Higgsino Dark Matter from Solar Capture}",
    eprint = "2609.02775",
    archivePrefix = "arXiv",
    primaryClass = "hep-ph",
    month = "9",
    year = "2026",
    journal = ""
}

@article{HCMDM,
    author = "Smirnov, Juri and Griffith, Spencer and Beacom, John F.",
    title = "{Inelastic Signatures of Electroweak Dark Matter}",
    eprint = "2609.04144",
    archivePrefix = "arXiv",
    primaryClass = "hep-ph",
    month = "9",
    year = "2026",
    journal = ""
}

@article{RSSX,
    author = "Rodd, Nicholas L. and Safdi, Benjamin R. and Slatyer, Tracy R. and Xu, Weishuang Linda",
    title = "{Confronting the Higgsino Interpretation of the LZ Event with the High-Energy Sideband}",
    eprint = "2609.04175",
    archivePrefix = "arXiv",
    primaryClass = "hep-ph",
    month = "9",
    year = "2026",
    journal = ""
}

@article{Nomura,
    author = "Nomura, Yasunori",
    title = "{Dark Matter as the $Z_2$ Partner of the Standard Model Higgs Boson}",
    eprint = "2609.02505",
    archivePrefix = "arXiv",
    primaryClass = "hep-ph",
    reportNumber = "RIKEN-iTHEMS-Report-26",
    month = "9",
    year = "2026",
    journal = ""
}

@article{WZZ,
    author = "Wu, Lei and Zhang, Yang and Zhu, Bin",
    title = "{TeV Higgsino Dark Matter from LZ Nuclear Recoil to Fermi-LAT Gamma Rays}",
    eprint = "2609.01590",
    archivePrefix = "arXiv",
    primaryClass = "hep-ph",
    month = "9",
    year = "2026",
    journal = ""
}

@article{Higgsino,
    author = "Freese, Katherine and Theodosopoulos, Dionysios P.",
    title = "{Higgsino Dark Matter Interpretation of the LUX-ZEPLIN 248 keV Nuclear-Recoil Event}",
    eprint = "2609.01583",
    archivePrefix = "arXiv",
    primaryClass = "hep-ph",
    month = "9",
    year = "2026",
    journal = ""
}

@article{DiMauro,
    author = "Di Mauro, Mattia",
    title = "{Dark Matter at the Kinematic Edge: Interpreting the 248 keV LZ Nuclear-Recoil Candidate}",
    eprint = "2609.02608",
    archivePrefix = "arXiv",
    primaryClass = "hep-ph",
    month = "9",
    year = "2026",
    journal = ""
}

@article{SYY,
    author = "Su, Liangliang and Yang, Jin Min and Yang, Wen-Na",
    title = "{Inelastic Dark Matter Signature at High Recoil Energy in LUX-ZEPLIN and CRESST}",
    eprint = "2609.01475",
    archivePrefix = "arXiv",
    primaryClass = "hep-ph",
    month = "9",
    year = "2026",
    journal = ""
}

@article{Vis,
    author = "Visinelli, Luca",
    title = "{A Peccei--Quinn Origin for Inelastic Electroweak Dark Matter after LUX-ZEPLIN}",
    eprint = "2609.02807",
    archivePrefix = "arXiv",
    primaryClass = "hep-ph",
    month = "9",
    year = "2026",
    journal = ""
}

@article{DP,
    author = "Yamashita, Kimiko",
    title = "{Inelastic Dark Photon Dark Matter for the LUX-ZEPLIN High-Recoil Event and the Galactic Halo Gamma-Ray Excess}",
    eprint = "2609.02868",
    archivePrefix = "arXiv",
    primaryClass = "hep-ph",
    month = "9",
    year = "2026",
    journal = ""
}

@article{DuWang,
    author = "Du, Xiaokang and Wang, Fei",
    title = "{TeV Higgsino Interpretation of the LZ High-Recoil Event with Intermediate-Scale Electroweak Gauginos}",
    eprint = "2609.04163",
    archivePrefix = "arXiv",
    primaryClass = "hep-ph",
    month = "9",
    year = "2026",
    journal = ""
}

@article{Yin,
    author = "Yin, Wen",
    title = "{A PQ-Symmetric High-Scale SUSY Interpretation of the LZ High-Energy Recoil}",
    eprint = "2609.01892",
    archivePrefix = "arXiv",
    primaryClass = "hep-ph",
    month = "9",
    year = "2026",
    journal = ""
}

@article{Singlino,
    author = "Chattopadhyay, Utpal and Das, Debottam and Puri, Rahul and Roy, Joydeep",
    title = "{Sub-TeV Singlino Dark Matter in light from Sagittarius A$^\ast$ and LUX-ZEPLIN Nuclear-Recoil Event}",
    eprint = "2609.02994",
    archivePrefix = "arXiv",
    primaryClass = "hep-ph",
    month = "9",
    year = "2026",
    journal = ""
}

@article{CKPT,
    author = "Cohen, Timothy and Kearney, John and Pierce, Aaron and Tucker-Smith, David",
    title = "{Singlet-Doublet Dark Matter}",
    eprint = "1109.2604",
    archivePrefix = "arXiv",
    primaryClass = "hep-ph",
    reportNumber = "MCTP-11-33, SLAC-PUB-14584",
    doi = "10.1103/PhysRevD.85.075003",
    journal = "Phys. Rev. D",
    volume = "85",
    pages = "075003",
    year = "2012"
}

@article{SD1,
    author = "Konar, Partha and Mukherjee, Ananya and Saha, Abhijit Kumar and Show, Sudipta",
    title = "{Linking pseudo-Dirac dark matter to radiative neutrino masses in a singlet-doublet scenario}",
    eprint = "2001.11325",
    archivePrefix = "arXiv",
    primaryClass = "hep-ph",
    doi = "10.1103/PhysRevD.102.015024",
    journal = "Phys. Rev. D",
    volume = "102",
    number = "1",
    pages = "015024",
    year = "2020"
}

@article{SD2,
    author = "Ghosh, Purusottam and Jeesun, Sk",
    title = "{Reviving sub-TeV $SU(2)_L$ lepton doublet dark matter}",
    eprint = "2306.12906",
    archivePrefix = "arXiv",
    primaryClass = "hep-ph",
    doi = "10.1140/epjc/s10052-023-12039-z",
    journal = "Eur. Phys. J. C",
    volume = "83",
    number = "9",
    pages = "880",
    year = "2023"
}

@article{SD3,
    author = "Bhattacharya, Subhaditya and Sahoo, Nirakar and Sahu, Narendra",
    title = "{Singlet-Doublet Fermionic Dark Matter, Neutrino Mass and Collider Signatures}",
    eprint = "1704.03417",
    archivePrefix = "arXiv",
    primaryClass = "hep-ph",
    doi = "10.1103/PhysRevD.96.035010",
    journal = "Phys. Rev. D",
    volume = "96",
    number = "3",
    pages = "035010",
    year = "2017"
}

@article{SDrev,
    author = "Bhattiprolu, Prudhvi N. and Petrosky, Evan and Pierce, Aaron",
    title = "{Singlet-doublet dark matter revisited}",
    eprint = "2505.11607",
    archivePrefix = "arXiv",
    primaryClass = "hep-ph",
    reportNumber = "LCTP-25-06",
    doi = "10.1103/pw56-v9z5",
    journal = "Phys. Rev. D",
    volume = "112",
    number = "3",
    pages = "035017",
    year = "2025"
}

@article{Catena,
    author = {Catena, Riccardo and Hellstr{\"o}m, Fredrik},
    title = "{New constraints on inelastic dark matter from IceCube}",
    eprint = "1808.08082",
    archivePrefix = "arXiv",
    primaryClass = "astro-ph.CO",
    doi = "10.1088/1475-7516/2018/10/039",
    journal = "JCAP",
    volume = "2018",
    number = "10",
    pages = "039",
    year = "2018"
}

@article{Gould,
    author = "Gould, Andrew",
    title = "{Resonant Enhancements in WIMP Capture by the Earth}",
    reportNumber = "SLAC-PUB-4226",
    doi = "10.1086/165653",
    journal = "Astrophys. J.",
    volume = "321",
    pages = "571",
    year = "1987"
}

@article{BS05,
    author = "Bahcall, John N. and Serenelli, Aldo M. and Basu, Sarbani",
    title = "{New solar opacities, abundances, helioseismology, and neutrino fluxes}",
    eprint = "astro-ph/0412440",
    archivePrefix = "arXiv",
    doi = "10.1086/428929",
    journal = "Astrophys. J. Lett.",
    volume = "621",
    pages = "L85--L88",
    year = "2005"
}

@article{IDM,
    author = "Tucker-Smith, David and Weiner, Neal",
    title = "{Inelastic dark matter}",
    eprint = "hep-ph/0101138",
    archivePrefix = "arXiv",
    reportNumber = "UCB-PTH-00-43, LBNL-47234, UW-PT-00-17",
    doi = "10.1103/PhysRevD.64.043502",
    journal = "Phys. Rev. D",
    volume = "64",
    pages = "043502",
    year = "2001"
}

@article{IDMstatus,
    author = "Tucker-Smith, David and Weiner, Neal",
    title = "{The Status of inelastic dark matter}",
    eprint = "hep-ph/0402065",
    archivePrefix = "arXiv",
    reportNumber = "UW-PT-04-01",
    doi = "10.1103/PhysRevD.72.063509",
    journal = "Phys. Rev. D",
    volume = "72",
    pages = "063509",
    year = "2005"
}

@article{MS06,
    author = "Mahbubani, Rakhi and Senatore, Leonardo",
    title = "{The Minimal model for dark matter and unification}",
    eprint = "hep-ph/0510064",
    archivePrefix = "arXiv",
    reportNumber = "MIT-CTP-3689, HUTP-05-A0044",
    doi = "10.1103/PhysRevD.73.043510",
    journal = "Phys. Rev. D",
    volume = "73",
    pages = "043510",
    year = "2006"
}

@article{DEramo,
    author = "D'Eramo, Francesco",
    title = "{Dark matter and Higgs boson physics}",
    eprint = "0705.4493",
    archivePrefix = "arXiv",
    primaryClass = "hep-ph",
    doi = "10.1103/PhysRevD.76.083522",
    journal = "Phys. Rev. D",
    volume = "76",
    pages = "083522",
    year = "2007"
}

@article{EFHPP,
    author = "Enberg, R. and Fox, P. J. and Hall, L. J. and Papaioannou, A. Y. and Papucci, M.",
    title = "{LHC and dark matter signals of improved naturalness}",
    eprint = "0706.0918",
    archivePrefix = "arXiv",
    primaryClass = "hep-ph",
    reportNumber = "LBNL-62748, UCB-PTH-07-10",
    doi = "10.1088/1126-6708/2007/11/014",
    journal = "JHEP",
    volume = "2007",
    number = "11",
    pages = "014",
    year = "2007"
}

@article{CMT,
    author = "Calibbi, Lorenzo and Mariotti, Alberto and Tziveloglou, Pantelis",
    title = "{Singlet-Doublet Model: Dark matter searches and LHC constraints}",
    eprint = "1505.03867",
    archivePrefix = "arXiv",
    primaryClass = "hep-ph",
    doi = "10.1007/JHEP10(2015)116",
    journal = "JHEP",
    volume = "2015",
    number = "10",
    pages = "116",
    year = "2015"
}

@article{BMMT,
    author = "Banerjee, Shankha and Matsumoto, Shigeki and Mukaida, Kyohei and Tsai, Yue-Lin Sming",
    title = "{WIMP Dark Matter in a Well-Tempered Regime: A case study on Singlet-Doublets Fermionic WIMP}",
    eprint = "1603.07387",
    archivePrefix = "arXiv",
    primaryClass = "hep-ph",
    reportNumber = "IPMU16-0039, LAPTH-013-16",
    doi = "10.1007/JHEP11(2016)070",
    journal = "JHEP",
    volume = "2016",
    number = "11",
    pages = "070",
    year = "2016"
}

@article{MDM,
    author = "Cirelli, Marco and Fornengo, Nicolao and Strumia, Alessandro",
    title = "{Minimal dark matter}",
    eprint = "hep-ph/0512090",
    archivePrefix = "arXiv",
    reportNumber = "DFTT40-2005, IFUP-TH-2005-34",
    doi = "10.1016/j.nuclphysb.2006.07.012",
    journal = "Nucl. Phys. B",
    volume = "753",
    pages = "178--194",
    year = "2006"
}

@article{MDMcosmo,
    author = "Cirelli, Marco and Strumia, Alessandro and Tamburini, Matteo",
    title = "{Cosmology and Astrophysics of Minimal Dark Matter}",
    eprint = "0706.4071",
    archivePrefix = "arXiv",
    primaryClass = "hep-ph",
    reportNumber = "IFUP-TH-2007-12, SACLAY-T07-052",
    doi = "10.1016/j.nuclphysb.2007.07.023",
    journal = "Nucl. Phys. B",
    volume = "787",
    pages = "152--175",
    year = "2007"
}

@article{ThomasWells,
    author = "Thomas, Scott D. and Wells, James D.",
    title = "{Phenomenology of Massive Vectorlike Doublet Leptons}",
    eprint = "hep-ph/9804359",
    archivePrefix = "arXiv",
    reportNumber = "SLAC-PUB-7799, SU-ITP-98-22",
    doi = "10.1103/PhysRevLett.81.34",
    journal = "Phys. Rev. Lett.",
    volume = "81",
    pages = "34--37",
    year = "1998"
}

@article{KrallReece,
    author = "Krall, Rebecca and Reece, Matthew",
    title = "{Last Electroweak WIMP Standing: Pseudo-Dirac Higgsino Status and Compact Stars as Future Probes}",
    eprint = "1705.04843",
    archivePrefix = "arXiv",
    primaryClass = "hep-ph",
    doi = "10.1088/1674-1137/42/4/043105",
    journal = "Chin. Phys. C",
    volume = "42",
    number = "4",
    pages = "043105",
    year = "2018"
}

@article{NagataShirai,
    author = "Nagata, Natsumi and Shirai, Satoshi",
    title = "{Higgsino Dark Matter in High-Scale Supersymmetry}",
    eprint = "1410.4549",
    archivePrefix = "arXiv",
    primaryClass = "hep-ph",
    reportNumber = "DESY-14-180, FTPI-MINN-14-37, IPMU14-0320",
    doi = "10.1007/JHEP01(2015)029",
    journal = "JHEP",
    volume = "2015",
    number = "01",
    pages = "029",
    year = "2015"
}

@article{GriestSeckel,
    author = "Griest, Kim and Seckel, David",
    title = "{Three exceptions in the calculation of relic abundances}",
    reportNumber = "CFPA-TH-90-001A, BA-90-79",
    doi = "10.1103/PhysRevD.43.3191",
    journal = "Phys. Rev. D",
    volume = "43",
    pages = "3191--3203",
    year = "1991"
}

@article{GondoloGelmini,
    author = "Gondolo, Paolo and Gelmini, Graciela",
    title = "{Cosmic abundances of stable particles: Improved analysis}",
    reportNumber = "UCLA-90-TEP-68",
    doi = "10.1016/0550-3213(91)90438-4",
    journal = "Nucl. Phys. B",
    volume = "360",
    pages = "145--179",
    year = "1991"
}

@article{Planck,
    author = "Aghanim, N. and others",
    collaboration = "Planck",
    title = "{Planck 2018 results. VI. Cosmological parameters}",
    eprint = "1807.06209",
    archivePrefix = "arXiv",
    primaryClass = "astro-ph.CO",
    doi = "10.1051/0004-6361/201833910",
    journal = "Astron. Astrophys.",
    volume = "641",
    pages = "A6",
    year = "2020",
    note = "[Erratum: Astron.Astrophys. 652, C4 (2021)]"
}

@article{PressSpergel,
    author = "Press, William H. and Spergel, David N.",
    editor = "Srednicki, M. A.",
    title = "{Capture by the sun of a galactic population of weakly interacting massive particles}",
    doi = "10.1086/163485",
    journal = "Astrophys. J.",
    volume = "296",
    pages = "679--684",
    year = "1985"
}

@article{GS87,
    author = "Griest, K. and Seckel, D.",
    title = "{Cosmic Asymmetry, Neutrinos and the Sun}",
    reportNumber = "CERN-TH-4505/86, SCIPP 86/60",
    doi = "10.1016/0550-3213(87)90293-8",
    journal = "Nucl. Phys. B",
    volume = "283",
    pages = "681--705",
    year = "1987",
    note = "[Erratum: Nucl.Phys.B 296, 1034--1036 (1988)]"
}

@article{JKG,
    author = "Jungman, Gerard and Kamionkowski, Marc and Griest, Kim",
    title = "{Supersymmetric dark matter}",
    eprint = "hep-ph/9506380",
    archivePrefix = "arXiv",
    reportNumber = "SU-4240-605, UCSD-PTH-95-02, IASSNS-HEP-95-14, CU-TP-677",
    doi = "10.1016/0370-1573(95)00058-5",
    journal = "Phys. Rept.",
    volume = "267",
    pages = "195--373",
    year = "1996"
}

@article{Nussinov,
    author = "Nussinov, Shmuel and Wang, Lian-Tao and Yavin, Itay",
    title = "{Capture of Inelastic Dark Matter in the Sun}",
    eprint = "0905.1333",
    archivePrefix = "arXiv",
    primaryClass = "hep-ph",
    doi = "10.1088/1475-7516/2009/08/037",
    journal = "JCAP",
    volume = "2009",
    number = "08",
    pages = "037",
    year = "2009"
}

@article{Menon,
    author = "Menon, Arjun and Morris, Rob and Pierce, Aaron and Weiner, Neal",
    title = "{Capture and Indirect Detection of Inelastic Dark Matter}",
    eprint = "0905.1847",
    archivePrefix = "arXiv",
    primaryClass = "hep-ph",
    reportNumber = "MCTP-09-14",
    doi = "10.1103/PhysRevD.82.015011",
    journal = "Phys. Rev. D",
    volume = "82",
    pages = "015011",
    year = "2010"
}

@article{IceCube,
    author = "Abbasi, R. and others",
    collaboration = "IceCube",
    title = "{Search for High-Energy Neutrinos From the Sun Using Ten Years of IceCube Data}",
    eprint = "2507.08457",
    archivePrefix = "arXiv",
    primaryClass = "hep-ex",
    month = "7",
    year = "2025",
    journal = ""
}

@article{IceCube3yr,
    author = "Aartsen, M. G. and others",
    collaboration = "IceCube",
    title = "{Search for annihilating dark matter in the Sun with 3 years of IceCube data}",
    eprint = "1612.05949",
    archivePrefix = "arXiv",
    primaryClass = "astro-ph.HE",
    doi = "10.1140/epjc/s10052-017-4689-9",
    journal = "Eur. Phys. J. C",
    volume = "77",
    number = "3",
    pages = "146",
    year = "2017",
    note = "[Erratum: Eur.Phys.J.C 79, 214 (2019)]"
}

@article{Helm,
    author = "Helm, Richard H.",
    title = "{Inelastic and Elastic Scattering of 187-Mev Electrons from Selected Even-Even Nuclei}",
    doi = "10.1103/PhysRev.104.1466",
    journal = "Phys. Rev.",
    volume = "104",
    pages = "1466--1475",
    year = "1956"
}

@article{LewinSmith,
    author = "Lewin, J. D. and Smith, P. F.",
    title = "{Review of mathematics, numerical factors, and corrections for dark matter experiments based on elastic nuclear recoil}",
    reportNumber = "RAL-TR-95-024",
    doi = "10.1016/S0927-6505(96)00047-3",
    journal = "Astropart. Phys.",
    volume = "6",
    pages = "87--112",
    year = "1996"
}

@article{SHM,
    author = "Baxter, D. and others",
    title = "{Recommended conventions for reporting results from direct dark matter searches}",
    eprint = "2105.00599",
    archivePrefix = "arXiv",
    primaryClass = "hep-ex",
    doi = "10.1140/epjc/s10052-021-09655-y",
    journal = "Eur. Phys. J. C",
    volume = "81",
    number = "10",
    pages = "907",
    year = "2021"
}

@article{RAVE,
    author = "Smith, Martin C. and others",
    title = "{The RAVE Survey: Constraining the Local Galactic Escape Speed}",
    eprint = "astro-ph/0611671",
    archivePrefix = "arXiv",
    doi = "10.1111/j.1365-2966.2007.11964.x",
    journal = "Mon. Not. Roy. Astron. Soc.",
    volume = "379",
    pages = "755--772",
    year = "2007"
}

@article{HisanoLoop,
    author = "Hisano, Junji and Ishiwata, Koji and Nagata, Natsumi and Takesako, Tomohiro",
    title = "{Direct Detection of Electroweak-Interacting Dark Matter}",
    eprint = "1104.0228",
    archivePrefix = "arXiv",
    primaryClass = "hep-ph",
    reportNumber = "IPMU-11-0046, ICRR-REPORT-583-2010-16, CALT-68-2824",
    doi = "10.1007/JHEP07(2011)005",
    journal = "JHEP",
    volume = "2011",
    number = "07",
    pages = "005",
    year = "2011"
}

@article{ChenHill,
    author = "Chen, Qing and Hill, Richard J.",
    title = "{Direct detection rate of heavy Higgsino-like and Wino-like dark matter}",
    eprint = "1912.07795",
    archivePrefix = "arXiv",
    primaryClass = "hep-ph",
    reportNumber = "FERMILAB-PUB-19-564-T",
    doi = "10.1016/j.physletb.2020.135364",
    journal = "Phys. Lett. B",
    volume = "804",
    pages = "135364",
    year = "2020"
}

@article{LZWIMP,
    author = "Aalbers, J. and others",
    collaboration = "LZ",
    title = "{Dark Matter Search Results from 4.2{\,}{\,}Tonne-Years of Exposure of the LUX-ZEPLIN (LZ) Experiment}",
    eprint = "2410.17036",
    archivePrefix = "arXiv",
    primaryClass = "hep-ex",
    reportNumber = "FERMILAB-PUB-24-0796-V",
    doi = "10.1103/4dyc-z8zf",
    journal = "Phys. Rev. Lett.",
    volume = "135",
    number = "1",
    pages = "011802",
    year = "2025"
}

@article{NeutrinoFog,
    author = "O'Hare, Ciaran A. J.",
    title = "{New Definition of the Neutrino Floor for Direct Dark Matter Searches}",
    eprint = "2109.03116",
    archivePrefix = "arXiv",
    primaryClass = "hep-ph",
    doi = "10.1103/PhysRevLett.127.251802",
    journal = "Phys. Rev. Lett.",
    volume = "127",
    number = "25",
    pages = "251802",
    year = "2021"
}

@article{CRESST,
    author = "Abdelhameed, A. H. and others",
    collaboration = "CRESST",
    title = "{First results from the CRESST-III low-mass dark matter program}",
    eprint = "1904.00498",
    archivePrefix = "arXiv",
    primaryClass = "astro-ph.CO",
    doi = "10.1103/PhysRevD.100.102002",
    journal = "Phys. Rev. D",
    volume = "100",
    number = "10",
    pages = "102002",
    year = "2019"
}

@article{CTA,
    author = "Acharyya, A. and others",
    collaboration = "CTA",
    title = "{Sensitivity of the Cherenkov Telescope Array to a dark matter signal from the Galactic centre}",
    eprint = "2007.16129",
    archivePrefix = "arXiv",
    primaryClass = "astro-ph.HE",
    doi = "10.1088/1475-7516/2021/01/057",
    journal = "JCAP",
    volume = "2021",
    number = "01",
    pages = "057",
    year = "2021"
}

@article{KraussWilczek,
    author = "Krauss, Lawrence M. and Wilczek, Frank",
    title = "{Discrete Gauge Symmetry in Continuum Theories}",
    reportNumber = "YCTP-P26-88, NSF-ITP-88-187",
    doi = "10.1103/PhysRevLett.62.1221",
    journal = "Phys. Rev. Lett.",
    volume = "62",
    pages = "1221",
    year = "1989"
}

@article{FanReece,
    author = "Fan, JiJi and Reece, Matthew",
    title = "{Higgsino Above the Sea of Fog}",
    eprint = "2609.01504",
    archivePrefix = "arXiv",
    primaryClass = "hep-ph",
    month = "9",
    year = "2026",
    journal = ""
}

@article{LouLu,
    author = "Lou, Yuanchao and Lu, Chih-Ting",
    title = "{Fermionic Dark Matter Absorption and the High-Energy Event in LUX-ZEPLIN}",
    eprint = "2609.01592",
    archivePrefix = "arXiv",
    primaryClass = "hep-ph",
    month = "9",
    year = "2026",
    journal = ""
}

@article{AtmNu,
    author = "Jeesun, Sk and Majumdar, Anirban",
    title = "{Atmospheric neutrino up-scattering explanation of LZ 2026 excess}",
    eprint = "2609.04185",
    archivePrefix = "arXiv",
    primaryClass = "hep-ph",
    month = "9",
    year = "2026",
    journal = ""
}

@article{HillSolonI,
    author = "Hill, Richard J. and Solon, Mikhail P.",
    title = "{Standard Model anatomy of WIMP dark matter direct detection I: weak-scale matching}",
    eprint = "1401.3339",
    archivePrefix = "arXiv",
    primaryClass = "hep-ph",
    reportNumber = "EFI-13-34",
    doi = "10.1103/PhysRevD.91.043504",
    journal = "Phys. Rev. D",
    volume = "91",
    pages = "043504",
    year = "2015"
}

@article{HillSolonII,
    author = "Hill, Richard J. and Solon, Mikhail P.",
    title = "{Standard Model anatomy of WIMP dark matter direct detection II: QCD analysis and hadronic matrix elements}",
    eprint = "1409.8290",
    archivePrefix = "arXiv",
    primaryClass = "hep-ph",
    reportNumber = "EFI-PREPRINT-14-25",
    doi = "10.1103/PhysRevD.91.043505",
    journal = "Phys. Rev. D",
    volume = "91",
    pages = "043505",
    year = "2015"
}

@article{FNW,
    author = "Fox, Patrick J. and Nelson, Ann E. and Weiner, Neal",
    title = "{Dirac gaugino masses and supersoft supersymmetry breaking}",
    eprint = "hep-ph/0206096",
    archivePrefix = "arXiv",
    reportNumber = "UW-PT-02-12, UW-02-12",
    doi = "10.1088/1126-6708/2002/08/035",
    journal = "JHEP",
    volume = "2002",
    number = "08",
    pages = "035",
    year = "2002"
}

@article{KPW,
    author = "Kribs, Graham D. and Poppitz, Erich and Weiner, Neal",
    title = "{Flavor in supersymmetry with an extended R-symmetry}",
    eprint = "0712.2039",
    archivePrefix = "arXiv",
    primaryClass = "hep-ph",
    doi = "10.1103/PhysRevD.78.055010",
    journal = "Phys. Rev. D",
    volume = "78",
    pages = "055010",
    year = "2008"
}

@article{Hsieh,
    author = "Hsieh, Ken",
    title = "{Pseudo-Dirac bino dark matter}",
    eprint = "0708.3970",
    archivePrefix = "arXiv",
    primaryClass = "hep-ph",
    reportNumber = "UMD-PP-07-004, MSU-HEP-07-08-28",
    doi = "10.1103/PhysRevD.77.015004",
    journal = "Phys. Rev. D",
    volume = "77",
    pages = "015004",
    year = "2008"
}

@article{McCabe,
    author = "McCabe, Christopher",
    title = "{Seasonal dark matter from the LUX-ZEPLIN high-energy event}",
    eprint = "2609.04181",
    archivePrefix = "arXiv",
    primaryClass = "hep-ph",
    month = "9",
    year = "2026",
    journal = ""
}

@article{DentNewstead,
    author = "Dent, James B. and Newstead, Jayden L.",
    title = "{Exothermic and Endothermic Inelastic Dark Matter Interpretations at LZ: Sideband Constraints and Future Prospects}",
    eprint = "2609.04673",
    archivePrefix = "arXiv",
    primaryClass = "hep-ph",
    month = "9",
    year = "2026",
    journal = ""
}

@article{Unwin,
    author = "Unwin, James",
    title = "{Axion Portal Dark Matter and the LUX-ZEPLIN High-Recoil Event}",
    eprint = "2609.04186",
    archivePrefix = "arXiv",
    primaryClass = "hep-ph",
    month = "9",
    year = "2026",
    journal = ""
}

@article{deLima,
    author = "de Lima, Carlos Henrique",
    title = "{Exothermic Dark Matter at LZ}",
    eprint = "2609.05204",
    archivePrefix = "arXiv",
    primaryClass = "hep-ph",
    month = "9",
    year = "2026",
    journal = ""
}

@article{GuLiTangXu,
    author = "Gu, Guanhua and Li, Lingfeng and Tang, Shao-Song and Xu, Yongheng",
    title = "{Inelastic from the Other Side: Xenon Excitation Signals in Light of the LZ High-Recoil Event}",
    eprint = "2609.05291",
    archivePrefix = "arXiv",
    primaryClass = "hep-ph",
    month = "9",
    year = "2026",
    journal = ""
}

@article{BGS,
    author = "Benakli, Karim and Goodsell, Mark D. and Staub, Florian",
    title = "{Dirac Gauginos and the 125 GeV Higgs}",
    eprint = "1211.0552",
    archivePrefix = "arXiv",
    primaryClass = "hep-ph",
    doi = "10.1007/JHEP06(2013)073",
    journal = "JHEP",
    volume = "2013",
    number = "06",
    pages = "073",
    year = "2013"
}

@article{DKKS,
    author = {Die{\ss}ner, Philip and Kalinowski, Jan and Kotlarski, Wojciech and St{\"o}ckinger, Dominik},
    title = "{Higgs boson mass and electroweak observables in the MRSSM}",
    eprint = "1410.4791",
    archivePrefix = "arXiv",
    primaryClass = "hep-ph",
    doi = "10.1007/JHEP12(2014)124",
    journal = "JHEP",
    volume = "2014",
    number = "12",
    pages = "124",
    year = "2014"
}

@article{HLMW,
    author = "Han, Tao and Liu, Hongkai and Mukhopadhyay, Satyanarayan and Wang, Xing",
    title = "{Dark Matter Blind Spots at One-Loop}",
    eprint = "1810.04679",
    archivePrefix = "arXiv",
    primaryClass = "hep-ph",
    doi = "10.1007/JHEP03(2019)080",
    journal = "JHEP",
    volume = "2019",
    number = "03",
    pages = "080",
    year = "2019"
}

@article{CRRS,
    author = "Chauhan, Bhavesh and Reno, Mary Hall and Rott, Carsten and Sarcevic, Ina",
    title = "{Neutrino constraints on inelastic dark matter captured in the Sun}",
    eprint = "2308.16134",
    archivePrefix = "arXiv",
    primaryClass = "hep-ph",
    doi = "10.1088/1475-7516/2024/01/030",
    journal = "JCAP",
    volume = "2024",
    number = "01",
    pages = "030",
    year = "2024"
}

@article{BorahSahoo2026,
    author = "Borah, Debasish and Sahoo, Sujit Kumar and Sahu, Narendra and Sharma, Shashwat",
    title = "{Inelastic Singlet-Doublet Fermion Dark Matter in light of the 248 keV LZ event}",
    eprint = "2609.07800",
    archivePrefix = "arXiv",
    primaryClass = "hep-ph",
    month = "9",
    year = "2026",
    journal = ""
}

@article{DiMauroShaikh2026,
    author = "Di Mauro, Mattia and Shaikh, Halim",
    title = "{Solar Capture Tests of Inelastic Dark Matter after the LZ High-Recoil Event}",
    eprint = "2609.06760",
    archivePrefix = "arXiv",
    primaryClass = "hep-ph",
    month = "9",
    year = "2026",
    journal = ""
}

@article{NguyenLindenHooper2026,
    author = "Nguyen, Thong T. Q. and Linden, Tim and Hooper, Dan",
    title = "{Solar Neutrino Constraints on Inelastic Dark Matter Scattering in Light of Recent LUX-ZEPLIN Observations}",
    eprint = "2609.11833",
    archivePrefix = "arXiv",
    primaryClass = "hep-ph",
    month = "9",
    year = "2026",
    journal = ""
}

@article{Langhoff2026,
    author = "Langhoff, Kevin",
    title = "{Heavy Higgsino Interpretation of the LZ Event}",
    eprint = "2609.09385",
    archivePrefix = "arXiv",
    primaryClass = "hep-ph",
    month = "9",
    year = "2026",
    journal = ""
}

@article{LeeRandall2026,
    author = "Lee, Vincent S. H. and Randall, Lisa",
    title = "{A Warped Extra Dimensional Candidate for the LZ 248 keV Event}",
    eprint = "2609.09136",
    archivePrefix = "arXiv",
    primaryClass = "hep-ph",
    reportNumber = "N3AS-26-020",
    month = "9",
    year = "2026",
    journal = ""
}

@article{ElahiSchwaller2026,
    author = "Elahi, Fatemeh and Schwaller, Pedro",
    title = "{A Vector-Like Lepton Interpretation of the High-Energy Nuclear Recoil Candidate in LUX-ZEPLIN}",
    eprint = "2609.08993",
    archivePrefix = "arXiv",
    primaryClass = "hep-ph",
    reportNumber = "MITP-26-043",
    month = "9",
    year = "2026",
    journal = ""
}

@article{KhanEtAl2026,
    author = "Khan, Imtiaz and Capozziello, Salvatore and Mustafa, G. and Atamurotov, Farruh and Abdujabbarov, Ahmadjon and Yuan, Chengxun",
    title = "{Nuclear interference versus dark sector excitation in the 248 keV LUX-ZEPLIN recoil candidate}",
    eprint = "2609.09230",
    archivePrefix = "arXiv",
    primaryClass = "hep-ph",
    month = "9",
    year = "2026",
    journal = ""
}

@article{He2026,
    author = "He, Yuxuan",
    title = "{Transition magnetic-dipole dark matter and the LZ230616 high-recoil candidate}",
    eprint = "2609.10453",
    archivePrefix = "arXiv",
    primaryClass = "hep-ph",
    month = "9",
    year = "2026",
    journal = ""
}

@article{FanEtAl2026,
    author = "Fan, Zi-Tong and He, Hong-Jian and Wang, Yu-Chen and Zhao, Yue",
    title = "{Inelastic Dark Matter and High-Energy Recoil Signatures in LZ}",
    eprint = "2609.10491",
    archivePrefix = "arXiv",
    primaryClass = "hep-ph",
    month = "9",
    year = "2026",
    journal = ""
}

@article{ChattarajEtAl2026,
    author = "Chattaraj, Ayan and Majumdar, Anirban and Papoulias, Dimitrios K. and Srivastava, Rahul",
    title = "{Can Elastic Neutrino Scattering Account for the LZ230616 Event?}",
    eprint = "2609.10504",
    archivePrefix = "arXiv",
    primaryClass = "hep-ph",
    month = "9",
    year = "2026",
    journal = ""
}

@article{BoseEtAl2026,
    author = "Bose, Debajit and others",
    title = "{Not so good $\nu$s for Higgsino dark matter as LZ excess: stringent limits from Super-Kamiokande and IceCube}",
    eprint = "2609.07807",
    archivePrefix = "arXiv",
    primaryClass = "hep-ph",
    month = "9",
    year = "2026",
    journal = ""
}

@article{WangXiao2026,
    author = "Wang, Lei and Xiao, Yang",
    title = "{The Inert Doublet Model of Dark Matter and the LUX-ZEPLIN High-Recoil Event}",
    eprint = "2609.06571",
    archivePrefix = "arXiv",
    primaryClass = "hep-ph",
    month = "9",
    year = "2026",
    journal = ""
}

@article{DasEtAl2026,
    author = "Das, Pritam and Karmakar, Biswajit and Mahapatra, Satyabrata and Paul, Partha Kumar",
    title = "{Inelastic Self-interacting Dark Matter and LUX-ZEPLIN 248 keV Event in a Dirac Modular Inverse Seesaw}",
    eprint = "2609.06825",
    archivePrefix = "arXiv",
    primaryClass = "hep-ph",
    month = "9",
    year = "2026",
    journal = ""
}

@article{AlhazmiEtAl2026,
    author = "Alhazmi, Haider and Kim, Doojin and Kong, Kyoungchul and Park, Jong-Chul and Shin, Seodong",
    title = "{High-Energy Nuclear Recoils from Boosted Dark Matter for the LZ 248-keV Event: Beyond the Halo-Dependent High-Velocity Tail}",
    eprint = "2609.06890",
    archivePrefix = "arXiv",
    primaryClass = "hep-ph",
    month = "9",
    year = "2026",
    journal = ""
}

@article{BisalCaoLi2026,
    author = "Bisal, Subhadip and Cao, Junjie and Li, Fei",
    title = "{Higgsino Dark Matter Interpretation of the LZ High-Recoil Event in the GNMSSM with TeV-Scale Gauginos}",
    eprint = "2609.07811",
    archivePrefix = "arXiv",
    primaryClass = "hep-ph",
    month = "9",
    year = "2026",
    journal = ""
}

@article{HMLee2026,
    author = "Lee, Hyun Min",
    title = "{Inelastic dark matter and baryon flavor symmetry in light of LUX-ZEPLIN (LZ) experiment}",
    eprint = "2609.06171",
    archivePrefix = "arXiv",
    primaryClass = "hep-ph",
    month = "9",
    year = "2026",
    journal = ""
}

@article{SaitoDT,
    author = "Saito, Masahiko and Sawada, Ryu and Terashi, Koji and Asai, Shoji",
    title = "{Discovery reach for wino and higgsino dark matter with a disappearing track signature at a 100 TeV $pp$ collider}",
    eprint = "1901.02987",
    archivePrefix = "arXiv",
    primaryClass = "hep-ph",
    doi = "10.1140/epjc/s10052-019-6974-2",
    journal = "Eur. Phys. J. C",
    volume = "79",
    pages = "469",
    year = "2019"
}

@article{HanMuC,
    author = "Han, Tao and Liu, Zhen and Wang, Lian-Tao and Wang, Xing",
    title = "{WIMPs at High Energy Muon Colliders}",
    eprint = "2009.11287",
    archivePrefix = "arXiv",
    primaryClass = "hep-ph",
    doi = "10.1103/PhysRevD.103.075004",
    journal = "Phys. Rev. D",
    volume = "103",
    pages = "075004",
    year = "2021"
}

@article{Ma2001,
    author = "Ma, Ernest",
    title = "{Naturally small seesaw neutrino mass with no new physics beyond the TeV scale}",
    journal = "Phys. Rev. Lett.",
    volume = "86",
    pages = "2502--2504",
    year = "2001",
    eprint = "hep-ph/0011121",
    archivePrefix = "arXiv"
}

@article{DavidsonLogan,
    author = "Davidson, Shainen M. and Logan, Heather E.",
    title = "{Dirac neutrinos from a second Higgs doublet}",
    journal = "Phys. Rev. D",
    volume = "80",
    pages = "095008",
    year = "2009",
    eprint = "0906.3335",
    archivePrefix = "arXiv"
}

@article{DiracNMSSM,
    author = "Lu, Xiaochuan and Murayama, Hitoshi and Ruderman, Joshua T. and Tobioka, Kohsaku",
    title = "{A Natural Higgs Mass in Supersymmetry from Nondecoupling Effects}",
    journal = "Phys. Rev. Lett.",
    volume = "112",
    pages = "191803",
    year = "2014",
    eprint = "1308.0792",
    archivePrefix = "arXiv"
}

@article{BHK,
    author = "Buckley, Matthew R. and Hooper, Dan and Kumar, Jason",
    title = "{Phenomenology of Dirac Neutralino Dark Matter}",
    journal = "Phys. Rev. D",
    volume = "88",
    pages = "063532",
    year = "2013",
    eprint = "1307.3561",
    archivePrefix = "arXiv"
}

@article{EFOS,
    author = "Ellis, John R. and Falk, Toby and Olive, Keith A. and Srednicki, Mark",
    title = "{Calculations of neutralino-stau coannihilation channels and the cosmologically relevant region of MSSM parameter space}",
    journal = "Astropart. Phys.",
    volume = "13",
    pages = "181--213",
    year = "2000",
    eprint = "hep-ph/9905481",
    archivePrefix = "arXiv"
}

@article{AHS,
    author = "Arkani-Hamed, Nima and Schmaltz, Martin",
    title = "{Hierarchies without symmetries from extra dimensions}",
    eprint = "hep-ph/9903417",
    archivePrefix = "arXiv",
    reportNumber = "SLAC-PUB-8082",
    doi = "10.1103/PhysRevD.61.033005",
    journal = "Phys. Rev. D",
    volume = "61",
    pages = "033005",
    year = "2000"
}

@article{GN,
    author = "Grossman, Yuval and Neubert, Matthias",
    title = "{Neutrino masses and mixings in nonfactorizable geometry}",
    eprint = "hep-ph/9912408",
    archivePrefix = "arXiv",
    reportNumber = "CLNS-99-1656, SLAC-PUB-8330",
    doi = "10.1016/S0370-2693(00)00054-X",
    journal = "Phys. Lett. B",
    volume = "474",
    pages = "361--371",
    year = "2000"
}

@article{GP,
    author = "Gherghetta, Tony and Pomarol, Alex",
    title = "{Bulk fields and supersymmetry in a slice of AdS}",
    eprint = "hep-ph/0003129",
    archivePrefix = "arXiv",
    reportNumber = "CERN-TH-2000-081, UNIL-IPT-00-06",
    doi = "10.1016/S0550-3213(00)00392-8",
    journal = "Nucl. Phys. B",
    volume = "586",
    pages = "141--162",
    year = "2000"
}

@article{DKYY,
    author = "Desai, Niral and Kilic, Can and Yang, Yuan-Pao and Youn, Taewook",
    title = "{Suppressed flavor violation in Lepton Flavored Dark Matter from an extra dimension}",
    eprint = "2001.00720",
    archivePrefix = "arXiv",
    primaryClass = "hep-ph",
    reportNumber = "UTTG 14-2019",
    doi = "10.1103/PhysRevD.101.075043",
    journal = "Phys. Rev. D",
    volume = "101",
    pages = "075043",
    year = "2020"
}

@article{GKMY,
    author = "Gignac, Matthew and Kilic, Can and Mahbubani, Rakhi and Youn, Taewook",
    title = "{Optimizing pixel tracklet searches for shorter lifetimes}",
    eprint = "2211.06949",
    archivePrefix = "arXiv",
    primaryClass = "hep-ph",
    reportNumber = "UTWI-14-2022, RBI-ThPhys-2022-41",
    doi = "10.1007/JHEP03(2023)040",
    journal = "JHEP",
    volume = "2023",
    number = "03",
    pages = "040",
    year = "2023"
}

@article{MS99,
    author = "Mirabelli, Eugene A. and Schmaltz, Martin",
    title = "{Yukawa hierarchies from split fermions in extra dimensions}",
    eprint = "hep-ph/9912265",
    archivePrefix = "arXiv",
    reportNumber = "SLAC-PUB-8309",
    doi = "10.1103/PhysRevD.61.113011",
    journal = "Phys. Rev. D",
    volume = "61",
    pages = "113011",
    year = "2000"
}

@article{KT00,
    author = "Kaplan, David Elazzar and Tait, Timothy M. P.",
    title = "{Supersymmetry breaking, fermion masses and a small extra dimension}",
    eprint = "hep-ph/0004200",
    archivePrefix = "arXiv",
    reportNumber = "ANL-HEP-PR-00-043, EFI-2000-12",
    doi = "10.1088/1126-6708/2000/06/020",
    journal = "JHEP",
    volume = "2000",
    number = "06",
    pages = "020",
    year = "2000"
}

@article{KT01,
    author = "Kaplan, David Elazzar and Tait, Timothy M. P.",
    title = "{New tools for fermion masses from extra dimensions}",
    eprint = "hep-ph/0110126",
    archivePrefix = "arXiv",
    reportNumber = "SLAC-PUB-9021, ANL-HEP-PR-01-081, EFI-01-44",
    doi = "10.1088/1126-6708/2001/11/051",
    journal = "JHEP",
    volume = "2001",
    number = "11",
    pages = "051",
    year = "2001"
}

@article{BCH16,
    author = "Blennow, Mattias and Clementz, Stefan and Herrero-Garcia, Juan",
    title = "{Pinning down inelastic dark matter in the Sun and in direct detection}",
    eprint = "1512.03317",
    archivePrefix = "arXiv",
    primaryClass = "hep-ph",
    doi = "10.1088/1475-7516/2016/04/004",
    journal = "JCAP",
    volume = "2016",
    number = "04",
    pages = "004",
    year = "2016"
}

@article{BCH18,
    author = "Blennow, Mattias and Clementz, Stefan and Herrero-Garcia, Juan",
    title = "{The distribution of inelastic dark matter in the Sun}",
    eprint = "1802.06880",
    archivePrefix = "arXiv",
    primaryClass = "hep-ph",
    doi = "10.1140/epjc/s10052-018-5863-4",
    journal = "Eur. Phys. J. C",
    volume = "78",
    number = "5",
    pages = "386",
    year = "2018",
    note = "[Erratum: Eur.Phys.J.C 79, 407 (2019)]"
}

@article{GouldSun,
    author = "Gould, Andrew",
    title = "{WIMP Distribution in and Evaporation From the Sun}",
    doi = "10.1086/165652",
    journal = "Astrophys. J.",
    volume = "321",
    pages = "560",
    year = "1987"
}

@article{KMSS21,
    author = "Konar, Partha and Mukherjee, Ananya and Saha, Abhijit Kumar and Show, Sudipta",
    title = "{A dark clue to seesaw and leptogenesis in a pseudo-Dirac singlet doublet scenario with (non)standard cosmology}",
    eprint = "2007.15608",
    archivePrefix = "arXiv",
    primaryClass = "hep-ph",
    doi = "10.1007/JHEP03(2021)044",
    journal = "JHEP",
    volume = "2021",
    number = "03",
    pages = "044",
    year = "2021"
}

@article{GKSS22,
    author = "Ghosh, Purusottam and Konar, Partha and Saha, Abhijit Kumar and Show, Sudipta",
    title = "{Self-interacting freeze-in dark matter in a singlet doublet scenario}",
    eprint = "2112.09057",
    archivePrefix = "arXiv",
    primaryClass = "hep-ph",
    doi = "10.1088/1475-7516/2022/10/017",
    journal = "JCAP",
    volume = "2022",
    number = "10",
    pages = "017",
    year = "2022"
}
